\documentclass[sigconf]{acmart}

\usepackage{mathtools}
\usepackage{tikz}
\usepackage{tikz}
\usepackage{listings}
\usepackage{comment}
\usepackage{enumerate}
\usepackage{color}
\usepackage{relsize}
\usepackage{booktabs}
\usepackage{graphics}
\usepackage[toc,page]{appendix} 

\newcommand{\subfigwidth}{0.235}
\newcommand{\subfigwidthThree}{0.29}
\newcommand{\bigfigwidth}{0.8\linewidth}

\usepackage{halloweenmath}
\usepackage{subcaption}
\usepackage{amsthm}
\usepackage{amsfonts}
\usepackage[ruled,linesnumbered]{algorithm2e}
\newcommand*{\defeq}{\stackrel{\text{def}}{=}}

\usepackage[ruled,linesnumbered]{algorithm2e}
\let\oldnl\nl
\newcommand{\nonl}{\renewcommand{\nl}{\let\nl\oldnl}}
\SetKwInput{KwData}{Input}
\SetKwInput{KwResult}{Output}
\SetKwComment{Comment}{/* }{ */}
\DontPrintSemicolon

\SetCommentSty{mycommfont}

\usepackage{xcolor}
\usepackage{url}
\usepackage{booktabs}

\newtheorem{definition}{Definition}
\newtheorem{lemma}{Lemma}
\usepackage{mathtools}

\definecolor{codegreen}{rgb}{0,0.6,0}
\definecolor{codegray}{rgb}{0.5,0.5,0.5}
\definecolor{codepurple}{rgb}{0.58,0,0.82}
\definecolor{backcolour}{rgb}{0.95,0.95,0.92}
\usepackage{listings}
\lstdefinestyle{mystyle}{
    backgroundcolor=\color{backcolour},   
    commentstyle=\color{codegreen},
    keywordstyle=\color{magenta},
    numberstyle=\tiny\color{codegray},
    stringstyle=\color{codepurple},
    basicstyle=\ttfamily\footnotesize,
    breakatwhitespace=false,         
    breaklines=true,                 
    captionpos=b,                    
    keepspaces=true,                 
    numbers=left,                    
    numbersep=5pt,                  
    showspaces=false,                
    showstringspaces=false,
    showtabs=false,                  
    tabsize=2
}

\theoremstyle{definition}

\usepackage[utf8]{inputenc}

\usepackage{seqsplit}

\begin{document}

\title{Order-Preserving Database Encryption with Secret Sharing}
\author{Dongfang Zhao}
\affiliation{%
   \institution{University of Nevada, Reno}
  \country{United States}
}

\begin{abstract}
The order-preserving encryption (OPE) problem was initially formulated by the database community in 2004 soon after the paradigm database-as-a-service (DaaS) was coined in 2002.
Over the past two decades, OPE has drawn tremendous research interest from communities of databases, cryptography, and security;
we have witnessed significant advances in OPE schemes both theoretically and systematically.
All existing OPE schemes assume that the outsourced database is modeled as a single semi-honest adversary who should learn nothing more than the order information of plaintext messages up to a negligible probability.
This paper addresses the OPE problem from a new perspective:
instead of modeling the outsourced database as a single semi-honest adversary,
we assume the outsourced database \textit{service} compromises a cluster of non-colluding servers,
which is a practical assumption as all major cloud vendors support multiple database instances deployed to exclusive sub-networks or even to distinct data centers.
This assumption allows us to design a new stateless OPE protocol,
namely order-preserving database encryption with secret sharing (ODES),
by employing secret-sharing schemes among those presumably non-colluding servers.
We will demonstrate that ODES guarantees the latest security level, namely IND-FAOCPA, and outperforms the state-of-the-art scheme by orders of magnitude. 
\end{abstract}

\settopmatter{printfolios=true}
\maketitle

\section{Introduction}

\subsection{Inception of Order-Preserving Encryption}

For two decades, we have witnessed the inception and prosperity of database as a service (DaaS) since the publication of the seminal paper~\cite{haci_icde02} in ICDE'02.
As of the writing of this paper, all major cloud computing vendors (e.g., Amazon Web Services, Google Cloud Platform, Microsoft Azure) support DaaS with pay-as-you-go business models,
which enables users to avoid the upfront cost of managing their in-house databases.
From the user's perspective, the DaaS can be thought of as an \textit{outsourced database} maintained by cloud computing vendors.
As with any outsourced service, the security of outsourced databases has been one of the top concerns for users:
the threat in an outsourced database comes not only from outside attackers but also inside adversaries, e.g., developers and administrators of the cloud computing vendor.
Among others, one avenue of research to address the above security issue is to encrypt the user's sensitive data before uploading them to the outsourced database.

One key challenge of DaaS lies in the management of those encrypted data,
such as building indexes,
because the index must be associated with the plaintext to speed up the query and modification requests from the user and yet all the server can learn about is the ciphertext.
To that end, in SIGMOD'04, Agrawal et al.~\cite{rag_sigmod04} proposed to encode the plaintext in the outsourced database while retaining the numerical order of the plaintext.
The paper demonstrated that it was possible to achieve the best of both worlds:
the so-called order-preservation encryption (OPE) can ensure both the confidentiality and the ordinal of the outsourced data.
The work quickly drew a lot of research interest from the database and the security/cryptography communities.

\subsection{Brief Timeline of OPE Security}

In EuroCrypt'09, Boldyreva et al.~\cite{abold_eurocrypt09} presented the first security definition of OPE.
Following the convention of cryptography, the definition is based on the canonical structure of an encryption scheme:
(i) the security goal is computational indistinguishability,
(ii) the threat model is to allow the adversary to obtain a polynomial number of ciphertexts of arbitrary plaintexts,
i.e., the so-called chosen-plaintext attack (CPA), and
(iii) a simulation-based reduction to prove that distinguishability is negligible.

Unfortunately, it was shown that it is impossible to achieve indistinguishability under the standard CPA attack because the CPA definition is overly strong and can be violated if the adversary can learn about the ordinal of the plaintexts.
The good news was that a new security notion,
namely \textit{indistinguishability under ordered chosen-plaintext attack} (IND-OCPA),
was proposed by~\cite{abold_eurocrypt09}.
IND-OCPA is as strong as IND-CPA except for allowing the adversary to only learn about the ordinal of the plaintext.

Not long after IND-OCPA was proposed, 
Popa et al.~\cite{rpopa_sp13} in Oakland'13 pointed out that IND-OCPA is insufficient for the well-known frequency attack.
The frequency attack is due to the deterministic ciphertexts,
which are known to be insecure under the conventional CPA attack as well.

A stronger security notion was then proposed in CCS'15,
namely \textit{indistinguishability under frequency-analyzing and ordered chosen-plaintext attack} (IND-FAOCPA)~\cite{fker_ccs15}.
Multiple subsequent schemes claimed to meet IND-FAOCPA,
such as~\cite{droche_ccs16}.
As of the writing of this paper, IND-FAOCPA remains the strongest security notion for order-preserving encryption schemes.

\subsection{System Research of OPE Schemes}

While the cryptography and security communities spent tremendous effort in properly defining and proving security from a theoretical perspective,
the database and system communities are equally interested in other metrics such as performance and costs.
Many leading cloud vendors are now supporting encrypted database services,
such as Microsoft Azure~\cite{panto_sigmod20} as reported in SIGMOD'20.

An evaluation paper~\cite{dboga_vldb19} in VLDB'19 summarized the pros and cons of major OPE schemes as of 2019.
The metrics include encryption complexity, comparison complexity, ciphertext size, I/O cost, and communication cost.
There was no clear winner based on the reported numbers in the paper.

In VLDB'21, Li et al.~\cite{dli_vldb21} presented a new frequency-hiding OPE scheme with a 128-bit AES (OPEA).
OPEA outperforms existing OPE schemes in almost all aspects:
\begin{itemize}
    \item OPEA is IND-FAOCPA secure;
    \item OPEA incurs a constant number of interactions between clients and servers;
    \item OPEA completes both insertion and query requests significantly faster than the counterparts;
    \item OPEA incurs $\mathcal{O}(N)$ client storage space, where $N$ denotes the number of distinct plaintexts.
\end{itemize}
The only limitation of OPEA lies in the client storage:
as a stateful scheme, OPEA requires the client to maintain a local table to keep track of the plaintext orders.
Although duplicate plaintexts only need to be stored one time,
counterparts (e.g., POPE~\cite{droche_ccs16}) could take constant client storage.
(However, POPE suffers the problem of possible incomparable elements)

\subsection{Motivation and Challenges of This Work}

The goal of this work is to eliminate the shortcoming of OPEA~\cite{dli_vldb21} while retaining its advantages compared with existing solutions.
This implies that we want to achieve both strong security levels and high performance.
By high performance, we mean lower processing time,
which is contributed by computational time, communication time, and I/O time.
Clearly, the $\mathcal{O}(N)$ client storage of OPEA is a performance bottleneck.

On the other hand, achieving both strong security guarantees and high performance is very challenging with currently available cryptographic primitives and our conventional wisdom of system optimization.
As discussed above, client storage is necessitated by the stateful coordination between the client and the server,
because a stateful mechanism is believed to achieve higher efficiency for maintaining the tuple orders.
As of the writing of this paper, 
we are only aware of one work~\cite{nshen_prdc21} taking a stateless approach;
but no experimental results were reported.
In addition, 128-AES is widely believed to be one of the most efficient and secure symmetric encryption schemes nowadays;
therefore, it is unlikely to improve the performance of OPEA by upgrading or optimizing the cryptographic subsystems.
Something more drastic is in need should we aim to further improve the performance without trading off the security level.

\subsection{Proposed Solution}

This work proposes a new OPE scheme by employing secret-sharing primitives.
While secret-sharing can be thought of as an encryption scheme (in a broad sense)
its internal machinery works quite differently than single-node encryption.
Instead of placing the encrypted test on a single machine,
we now assume a cluster of machines that would not collude.
The ciphertext is now distributed into multiple database servers as secret shares and cannot be decrypted without the authorization of the data owner.
Therefore, confidentiality is achieved.

To achieve the ordinal of plaintexts,
the secret-sharing primitives should allow us to compare the corresponding plaintext values by asking the server not to share its local secret shares.
If we can achieve this comparison merely through some local computation of the secret shares,
the resulting scheme would be stateless and save us some I/O costs.

\begin{figure}[!t]
  \centering
  \includegraphics[width=\linewidth]{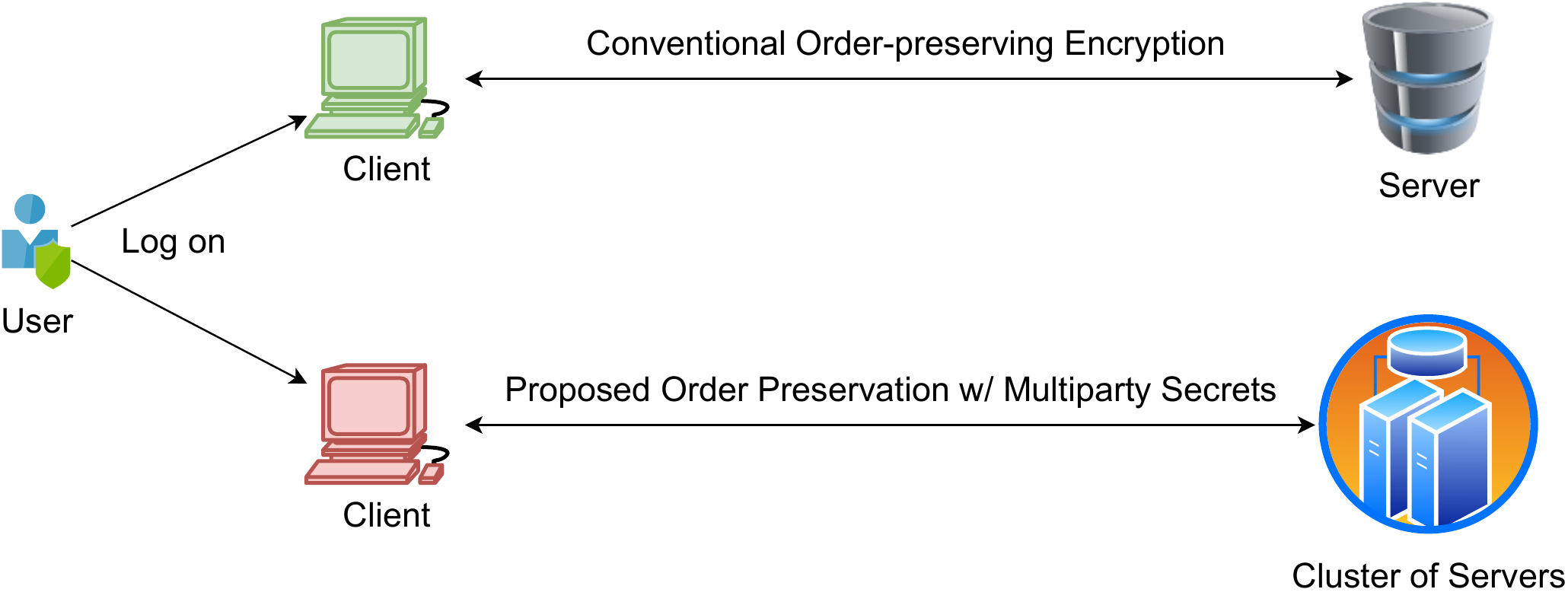}
  \caption{Proposed multiparty secrets vs. conventional single-node encryption.}
  \label{fig:intro}
\end{figure}

As we will demonstrate in the latter sections,
a specific type of secret-sharing scheme does allow us to achieve both confidentiality and ordinal of encoded data.
To make matters more concrete, 
Figure~\ref{fig:intro} illustrate the high-level difference between the conventional OPE schemes and the proposed scheme,
which we coin as \textit{ordered database encryption with secret-sharing} (ODES).

\subsection{Contributions}

In summary, this work makes the following technical contributions:
\begin{itemize}
    \item We propose the very first stateless order-preserving encryption scheme for outsourced databases with secret sharing, namely ordered database encryption with secret-sharing (ODES);
    \item We demonstrate that ODES guarantees a strong security level,
    i.e., IND-FAOCPA;
    \item We design various database protocols for leveraging the proposed ODES scheme; 
    \item We implement the ODES scheme and database protocols on top of SQLite databases; and
    \item We conduct a thorough evaluation of ODES by comparing it with state-of-the-art schemes on top of the TPC-H benchmark and three real-world applications on a 10-node cluster.
\end{itemize}

\section{Background and Related Work}

\subsection{Order-Preserving Encryption}

The concept of order-preserving encryption (OPE) was originally proposed in the database community~\cite{rag_sigmod04}.
The motivation is evident:
how could we achieve both the confidentiality and the ordinals of sensitive data in an outsourced database?
The confidentiality part is obvious and the ordinal requirement is also well justified:
it is very common for database systems to build indexes to speed up the query and insertion requests and being able to sort or order the outsourced data sets is essential to achieve this goal.

The early-state solution to achieve the dual goals is somewhat straightforward:
the plaintexts are encoded with the help of some statistical distribution such that the encoded values remain in the same order as the plaintext.
There are a few issues with this approach.
First, the encoded values are deterministic.
This means the encoding cannot be secure if the adversary can somehow obtain the encoding of some chosen plaintexts,
i.e., the so-called chosen-plaintext attack (CPA).
It can be argued that in outsourced databases we do not need security as strong as CPA,
but from a security point of view, a practical database system should always provide CPA security as the minimum~\cite{panto_sigmod20}.

We then run into a dilemma between CPA security and ordinal encoding:
CPA security implies the randomness of ciphertexts which cannot retain the order of plaintexts.
The solution is to introduce some function for the database to order the encrypted tuples without relying on the raw values of ciphertexts,
which is called \textit{order-revealing encryption} (ORE)~\cite{dbone_eurocrypt15}.
Accordingly, a new security notion was proposed to allow the adversary to learn about the orders of plaintexts,
resulting in the so-called indistinguishability under ordered chosen-plaintext attack (IND-OCPA).
Obviously, there are many options to calculate the order values but they can be categorized into two categories:
(i) a stateful scheme where the client and the server coordinate to maintain the order information of encrypted records in the database~\cite{rpopa_sp13}, and
(ii) a stateless scheme where the order information can be retrieved on the fly~\cite {nshen_prdc21}.
Most OPE works in the literature focus on the stateful approach;
as we will see in the latter sections, the proposed ODES is a stateless scheme.

It turned out that there were new issues for ORE and IND-OCPA:
Many IND-OCPA schemes~\cite{abold_eurocrypt09,rpopa_sp13} are vulnerable to attacks that leverage the access patterns of the queries.
To that end, a newer notion is defined,
the indistinguishability under frequency-analyzing ordered chosen-plaintext attack (IND-FAOCPA).
Multiple IND-FAOCPA schemes have been proposed in the literature,
such as~\cite{dli_vldb21,fker_ccs15,droche_ccs16}.
A relatively recent evaluation paper~\cite{dboga_vldb19} reports the performance of some of the most popular OPE schemes,
including what have not been mentioned in this paper:~\cite{nchen_fse16,klewi_ccs16,dcash_asiacrypt}.
As of the writing of this paper, the OPE scheme proposed in~\cite{dli_vldb21} achieves the best performance in almost all the metrics (e.g., client storage, query rounds),
and we will primarily compare the proposed ODES with the protocol proposed in~\cite{dli_vldb21}.

\subsection{Secret Sharing}

The idea of a secret sharing scheme (SSS) is straightforward:
a given plaintext $pt$ is converted into a set of encoded bytes $ct$'s such that only a specific \textit{subset} of $ct$'s can reconstruct the original $pt$.
The goal of SSS is to reduce the risk of disclosing the plaintext;
instead of compromising the holder of the plaintext,
the malicious adversary needs to compromise multiple entities before any of the shareholders detect the attack.
Even for weaker attacks where only semi-honest adversaries are assumed, 
dispersing the secret shares to more parties raises the bar of a successful eavesdropping attack.

The SSS can be tuned by the subset size. 
Formally, a $t$-out-of-$n$ threshold SSS (TSSS) is defined as follows.
\begin{definition}[TSSS]
A TSSS is comprised of two algorithms:
\begin{itemize}
    \item Share: a randomized algorithm that takes as input a plaintext $pt$ and returns a sequence $S = (s_1, \dots, s_n)$ of shares.
    \item Reconstruct: a deterministic algorithm that takes a set of at least $t$ shares and returns the plaintext.
\end{itemize}
The number $t$ is called the threshold of the TSSS.
Let $U$ of size $t$ be a subset of $n$ shares, $|U| \ge t$ and $U \subseteq \{s_1, \dots, s_n\}$,
we require that a TSSS holds the following property:
\[
Reconstruct(U) = pt.
\]
\end{definition}
As we will see in the next section~\S\ref{sec:provable}, the definition of TSSS leads to a slightly different security definition compared with the conventional encryption schemes.

The canonical example of TSSS is due to Shamir~\cite{sham_ccam79},
in which the secrets are revealed through a $(t-1)$-degree polynomial.
In essence, each share can reconstruct the coefficient of a specific degree of unknowns through the LaGrange polynomials.
In addition to Shamir's construction, 
other schemes exist.
Ito et al.~\cite{mito_ecj89} proposed the replicated secret-sharing scheme,
which is based on finite fields where each share is a vector.
One nice property of replicated secret-sharing is its linearity:
the addition and subtraction of local shares equal the addition and subtraction of the plaintext.
A simpler variant of replicated secret-sharing is \textit{additive secret sharing},
where each share is a scalar value and the threshold $t$ is set to $n$.
Indeed, this is a building block of our proposed ODES protocol;
we will see how the linearity of the additive secret-sharing allows us to preserve the orders of plaintexts in~\S\ref{sec:primitive}.

SSS has a tight connection with secure multiparty computation (MPC)~\cite{daran_latincrypt21},
which has a long history~\cite{ayao_focs82}.
The goal of MPC is more ambitious than SSS:
in addition to keeping the plaintext confidential,
we want to calculate an arbitrary function of the original plaintexts by touching on only the encoded data on multiple parties.
The original problem was solved by the so-called garbled circuits~\cite{ayao_focs82},
whose idea was pretty simple:
we can ask each party to encode the input with its private key,
shuffle the encrypted ciphertexts,
and then enumerate all the keys to decrypt the result.
Since we assume the encryption scheme is secure,
the only way that the result can be revealed is that the correct combination of private keys is applied to one of the garbled outputs.
This is indeed a feasible solution, at least theoretically;
in practice, the circuits may grow exponentially and result in efficiency issues.
There are many more efficient MPC solutions,
such as~\cite{dbeav_stoc90,mnaor_ec99,vkole_asiacrypt05,szahu_eurocrypt15}.

\subsection{Provable Security}\label{sec:provable}

When employing an encryption scheme in an application,
it is highly desirable to demonstrate its security provably.
Formally, we need to identify the following three important pieces for the provable security of a given encryption scheme:
security goal, threat model, and assumptions.
The security goal spells out the desired effect when the application is under attack;
the threat model articulates what an adversary can do with the attack,
such as what information of the plaintext/ciphertext can be collected and the resource/time limitation of the attack;
the assumption lists the presumed specifics of the subsystems or components of the cryptographic scheme,
which is usually an important building block for security proof, e.g., reduction.
The security goal and threat model are usually called \textit{security definition} collectively.

One well-accepted security definition with a good balance between efficiency and security is that the adversary can launch a \textit{chosen-plaintext attack} (CPA),
defined as follows.
\begin{definition}[Chosen-Plaintext Attack]
Given a security parameter $n$,
i.e., the bitstring length of the key,
an adversary can obtain up to $poly(n)$ of plaintext-ciphertext pairs $(m, c)$,
where $m$ is arbitrarily chosen by the adversary and $poly(\cdot)$ is a polynomial function on $n$.
With such information, the adversary tries to decrypt a $c'$ that is not included in the polynomial number of known ciphertexts.
\end{definition}

The polynomial requirement is only for practical reasons,
as we usually assume that the adversary should only be able to run a polynomial algorithm without unlimited resources.
Accordingly, we want to design encryption schemes that are \textit{CPA secure}: 
even if the adversary $\mathcal{A}$ can obtain those extra pieces of information, 
$\mathcal{A}$ should not be able to decode the ciphertext better than a random guess up to a very small probability.
To quantify the degree of this small probability,
\textit{negligible function} is defined as below.

\begin{definition}
A function $\mu(\cdot)$ is called negligible if for all polynomials $poly(n)$ the inequality $\mu(n) < \frac{1}{poly(n)}$ holds for sufficiently large $n$'s.
\end{definition}

For completeness, we list the following lemmas for negligible functions that will be used in later sections.
We state them without the proofs, 
which can be found in introductory cryptography or complexity theory texts.

\begin{lemma}[Summation of two negligible functions is a negligible function]
\label{thm:neg_sum}
Let $\mu_1(n)$ and $\mu_2(n)$ be both negligible functions.
Then $\mu(n)$ is a negligible function that is defined as $\mu(n) \defeq \mu_1(n) + \mu_2(n)$.
\end{lemma}

\begin{lemma}[Quotient of a polynomial function over an exponential function is a negligible function]\label{thm:neg_quotient}

$\frac{ploy(n)}{2^n}$ is a negligible function. 
That is,
$\exists N \in \mathbb{N},\; \forall n \geq N:\; \frac{ploy(n)}{2^n} < \frac{1}{poly(n)} $, where $\mathbb{N}$ denotes natural numbers. 
\end{lemma}

The canonical method to prove the security of a proposed encryption scheme,
such as IND-CPA,
is through \textit{reduction}~\cite{ylind_book17}.
Usually, breaking the scheme is \textit{reduced} to a hard mathematical problem,
which means that if an attack is possible for the scheme then the mathematical problem would be efficiently solved.
That is, the encryption scheme is not easier than the mathematical problem.
The scheme is modeled as a subroutine,
whose inputs are simulated such that the adversary cannot tell whether it is being involved in an attack or in a subroutine to help solve the math problem. 
Although forward proof is possible,
the more commonly used technique is the contradiction:
by assuming that the adversary could distinguish some designed experiments with a non-negligible advantage, 
the reduction would lead to a non-negligible probably to efficiently solve the hard mathematical problem that is believed to be intractable,
thus leading to a contradiction.

Unfortunately, in the context of order-preserving encryption, 
it has been proven that the conventional IND-CPA is impossible~\cite{abold_eurocrypt09}.
Therefore, the cryptography community proposed a relaxed notation called \textit{indistinguishability under ordered chosen-plaintext attack} (IND-OCPA)~\cite{abold_eurocrypt09}.
However, it was shown that~\cite{vbind_vldb18} effective attacks can be launched on IND-OCPA security caused by the access patterns.
The root cause of this issue lies in the deterministic ciphertext in early-stage order-preserving encryption schemes.
Indeed, it is well known that a deterministic encryption scheme can be impossibly secure against CPA attacks.
As a result, modern order-preserving encryption schemes are all randomized,
which implies that the ciphertexts are not directly comparable and necessities indirect comparison between ciphertexts.
Such indirection comparison is usually coined as \textit{order-revealing encryption} (ORE)~\cite{dbone_eurocrypt15} that generalizes the original notion of OPE.
In a more general sense, some so-called frequency-hiding order-preserving encryption schemes~\cite{fker_ccs15,droche_ccs16} were proposed.
Accordingly, a new security notion was proposed,
namely indistinguishability under frequency-analyzing ordered chosen-plaintext attack (IND-FAOCPA)~\cite{fker_ccs15}.
IND-FAOCPA is the latest security definition in this area and our proposed ODES scheme is IND-FAOCPA secure.

The above review of provable security assumes that the ciphertext is a single entity and does not consider the scenario where the ciphertext is a \textit{set},
which is the case for secret sharing.
The provable security of secret sharing takes a slightly different approach because of the additional assumption that not all shares will be accessible to the adversary per the definition.
While it is true that if we can prove the entire set of shares is secure then any subset of shares is also secure,
a more common approach to proving the semantic security of secret shares is through \textit{interchangeable libraries}~\cite{mrosu_book}.
The key idea is to model the scheme as a library with the input of either an $L$ or $R$ plaintext input,
and then the proof will show that the library with $L$ input eventually looks identical to the library with the $R$ input through a series of interchangeable operations.
We will see how this technique is used in~\S\ref{sec:security}

\section{Order-Preserving Database Encryption with Secret Sharing}

\subsection{Overview}

The intuitive idea behind the proposed secret-sharing-based order preservation is straightforward:
we leverage the multiplicity of a cluster of database servers such that no plaintext is leaked while maintaining the comparative order among the plaintexts.
That is, we somehow break the original plaintext into multiple shares,
each of which is allocated to a distinct server in a database cluster.
The nodes in the cluster are assumed to be non-colluding,
which can be implemented by deploying the servers into different sub-networks or different data centers.
The order of the plaintext can be retrieved and updated by an aggregation of local functions on each server.

\begin{figure}[!t]
  \centering
  \includegraphics[width=\linewidth]{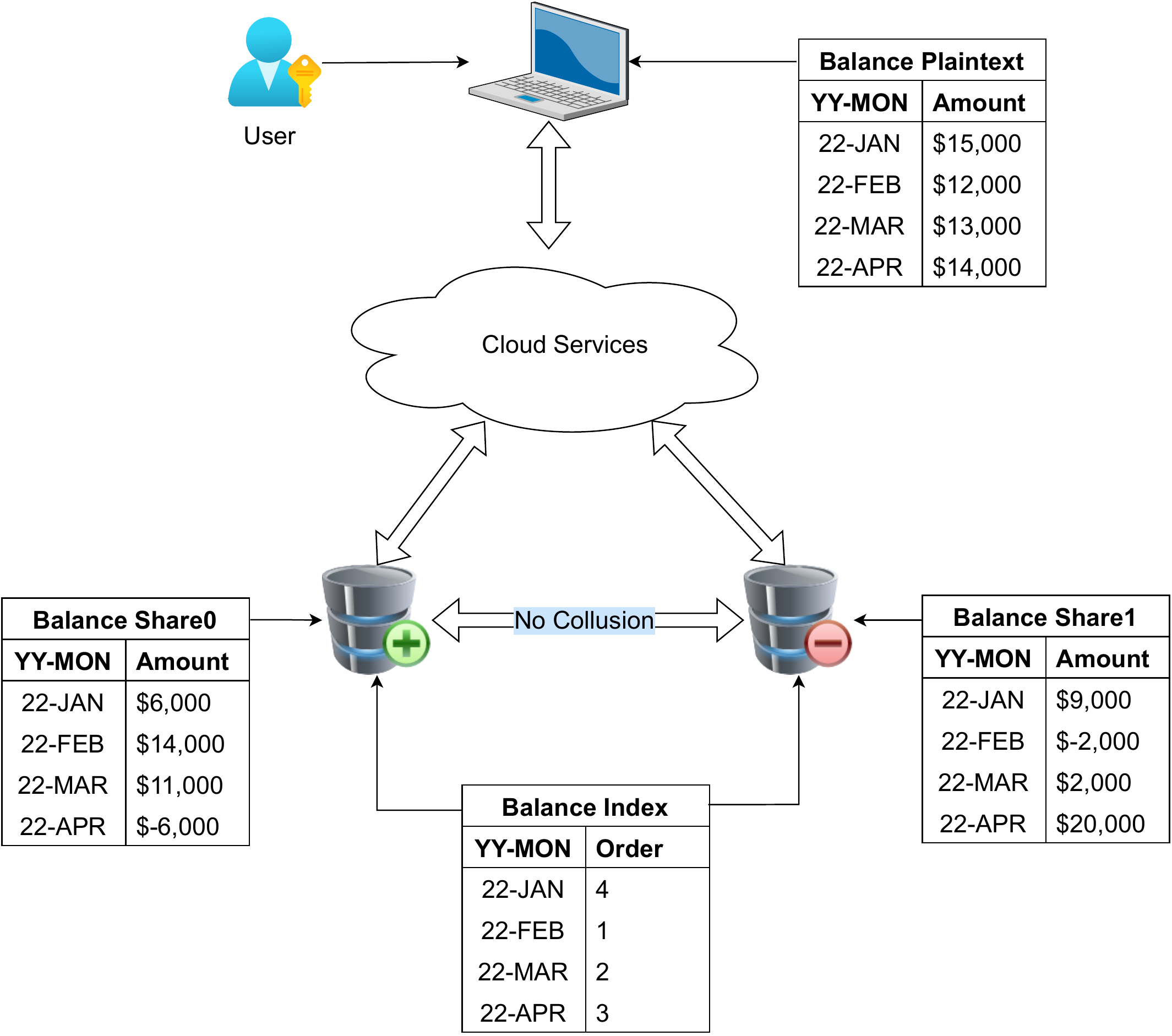}
  \caption{Example of two shares with order preservation.}
  \label{fig:odes_example}
\end{figure}

To make matters more concrete,
Figure~\ref{fig:odes_example} illustrate the idea in an oversimplified scenario where two servers are available to store encoded and ordered data.
Assume that the data owner wants to save the balance table into a remote database service.
The table is as simple as a key-value store with the year-month as the key and the U.S. dollar amount as the value for his business.

Each of the two database servers stores some seemingly random numbers that will help keep the real amounts confidential. 
We must be able to reconstruct the original plaintext from the shares stored by the servers because otherwise, the user would lose the data.
In this example, the original balance amount can be reconstructed by simply adding up the shares from distinct database servers.
It should be noted that the reconstruction should only happen on the client side as the outsourced databases are not fully trusted.
That is, the database servers are not supposed to share their local data.
We will formally define what we mean by ``fully trusted'' later;
but for now, we keep our discussion at a non-technical level.

Another important piece of information is the index metadata for keeping track of the order information of the plaintext.
The index table will be useful when a new record is inserted into the databases:
it allows us to do a binary search to locate the correct order of the newly inserted record in the database.
In addition, it would facilitate the effectiveness of order-related queries.

Here is another example for illustrating the plaintext comparison through two sets of secret shares.
Let's say the user recently obtained a new record for (22-MAY, \$10,000) and would like to know whether the balance is higher than the previous month, 22-APR.
Note that ODES is a stateless protocol, 
so the client cannot simply compare \$10,000 to (22-APR, \$14,000) since the latter does not exist after being secretly shared with the two database servers.
What the client would do is split \$11,000 into two random numbers,
say \$3,000 and \$8,000,
which are combined with the key and sent to the two database servers.
In our example of Fig.~\ref{fig:odes_example},
the plus server receives (22-MAY, \$3,000) record and the minus server receives (22-MAY, \$8,000) record.
Both servers then carry out \textit{local} computations of the 22-APR and 22-MAY records;
for example, on the plus server, it calculates the following delta ($\delta_0$, 3000-(-6000) = 9000).
Similarly, the minus server calculates ($\delta_1$, 8000-20000 = -12000).
Both servers then broadcast their local $\delta_j$ to all other servers.
Note that this sharing would not reveal any information other than the orders and is therefore allowed (as opposed to the secret share itself).
Now, each database server has obtained all the $\delta_j$'s and then applies an aggregation over them.
For example, the plus server calculates 9000+(-12000) = -3000 < 0,
which means the balance of 22-MAY is lower than 22-APR.
The server can also update the index metadata accordingly by conducting a binary search on the index file;
in this example, (22-MAY, 1) will be inserted into the index table and some existing orders will be incremented by one.

\subsection{Architecture}

\begin{figure}[!t]
  \centering
  \includegraphics[width=\linewidth]{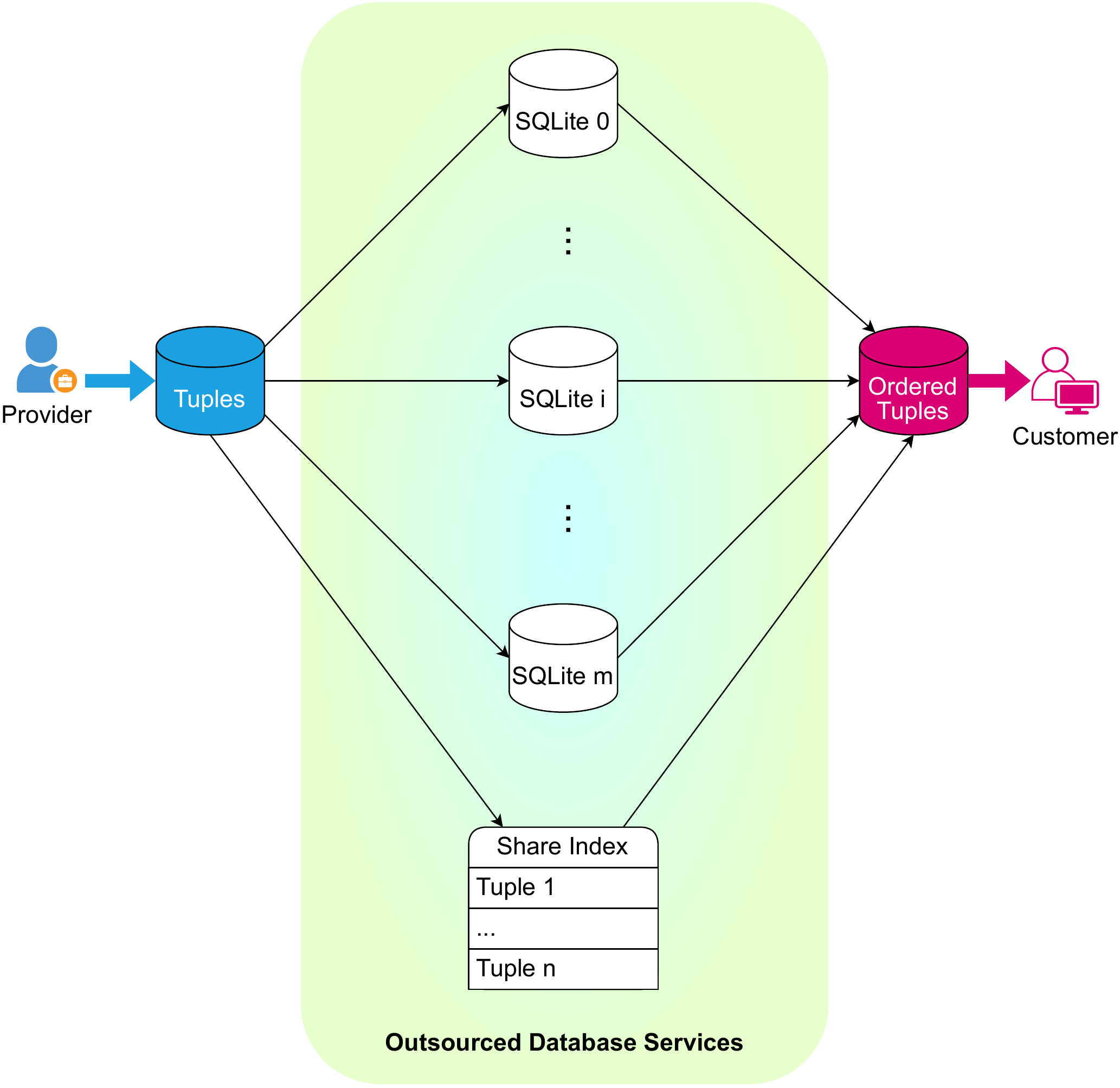}
  \caption{The proposed architecture of preserving tuple orders with secret shares in outsourced SQLite databases.}
  \label{fig:arch}
\end{figure}

As shown in Figure~\ref{fig:arch}, we envision a cluster of $m$ outsourced database servers that 
(i) do not share their local data and
(ii) can access an index for the orders of the secret shares.
Since our system prototype is implemented with SQLite,
we assume there are $m$ SQLite instances.
As long as at least one of the $m$ SQLite instances does not collude with others,
the outsourced data is secure,
which overcomes the so-called \textit{dishonest majority} problem in the literature.
This implies that more databases imply a higher security level;
however, this is at the cost of more computational and I/O overhead in the system.

The architecture also differentiates two different user roles:
a data provider (on the left) and a data customer (on the right).
The key difference is that the provider may modify the secret shares in the database cluster and possibly update the index metadata as well,
while the customer only makes read-only queries.
We will see that both modification and read-only queries will be facilitated by the share index in the proposed protocols.

\subsection{Primitives}\label{sec:primitive}

We are now ready to formally present the primitives of ODES to allow order preservation. 
These primitives will be used as building blocks in various protocols later (\S\ref{sec:protocol}).
We assume that the plaintext is a string of $l$ bits and there are overall $m$ database servers.

\paragraph{Share}
We start with the \textit{Share} function that splits a given plaintext into $m$ secret shares.
\begin{definition}[Share]
A Share() function is defined as
\[\displaystyle
\begin{split}
Share: \{0,1\}^l & \to \left(\{0,1\}^l\right)^m, \\
pt & \mapsto \{s_i\}, 0 \le i < m,
\end{split}
\]
where $s_i$ is a random number when $i \not= 0$ and $s_0$ is calculated as
\[\displaystyle
s_0 \coloneq \sum_{i=1}^{m-1} s_i.
\]
\end{definition}

\paragraph{Reconstruct}
The \textit{Reconstruct} function is the reverse of \textit{Share}.
\begin{definition}[Reconstruct]
A Reconstruct() function is defined as
\[\displaystyle
\begin{split}
Reconstruct: \left(\{0,1\}^l\right)^m & \to \{0,1\}^l, \\
\{s_0, \dots, s_{m-1}\} & \mapsto \sum_{i=0}^{m-1} s_i.
\end{split}
\]
\end{definition}

\paragraph{Compare}
A cluster of $m$ servers aims to compare two plaintexts by local computations of two sets of secret shares.
As the name suggests, a server is not allowed to disclose its local share.
Let $\displaystyle S^L \defeq \left\{s^L_0, \dots, s^L_{m-1}\right\}$ denote the secret shares of the first plaintext $L$ and $\displaystyle S^R \defeq \left\{s^R_0, \dots, s^R_{m-1}\right\}$ denote the secret shares of the second plaintext $R$.
\begin{definition}[Compare]
A Compare is a function:
\[\displaystyle
\begin{split}
Compare: \left(\{0, 1\}^l\right)^m \times \left(\{0, 1\}^l\right)^m & \to \{-1, 0, 1\},\\
(S^L, S^R) & \mapsto \begin{cases}
-1, & L < R \\
0, & L = R \\
1, & L > R
\end{cases},
\end{split}
\]
where Compare is assigned the same arithmetic sign as the following summation
\[\displaystyle
    \sum_{i=0}^{m-1} \left( s^L_i - s^R_i \right).
\]
\end{definition}
Note that the terms in the above equation are calculated by each database server independently.
The servers broadcast their local differences to the entire cluster such that each server can decide the ordinal information of the plaintexts.

\subsection{Protocols}\label{sec:protocol}

We present three ODES protocols in this section.
Various primitives will be used in the protocols.
To make the protocols self-contained, 
we include some high-level implementation details of the primitives;
when we do so, we make comments to the pseudocode to remind the readers that certain primitives are being called (e.g., Line 3 in Alg.~\ref{alg:odes_init}).
We follow the cryptographic convention to use $\coloneq$ to denote a deterministic assignment and $\gets$ to denote a uniformly sampled value from a randomized algorithm or distribution.

\subsubsection{Initialization}

When the system is deployed for the first time,
we assume that the data provider has an initial list of data sets that will be encoded and uploaded to the remote database service.
Several tasks must be completed before the proposed ODES system goes into operation,
including:
\begin{itemize}
    \item An initial index file is built by the data provider based on the local plaintexts;
    \item Each plaintext field is decomposed into a list of shares;
    \item The shares of the same plaintext should be randomized and distributed to distinct database servers.
\end{itemize}

All of those tasks are completed in the initialization phase,
as illustrated in Alg.~\ref{alg:odes_init}.
For the $i$-th record $r$, the client (i.e., the provider in Fig.~\ref{fig:arch}) splits it into $m$ pieces (Lines 3--6):
the last $m-1$ pieces $s_i[1:j]$ are simply random values and the first piece is calculated as the difference between $r$ and $\sum s_i[1:j]$.
To add more randomness to the encoded ciphertext,
we apply permutation to the $m$ shares as well (Line 7).
In Line 9, the client sends each of the $m$ shares to a distinct database server;
in Line 10, the server receives the share and inserts it into its local share.
After processing the record $r$, the index metadata is updated in Line 12.
After completing all the $n$ records,
the client broadcasts the index table to all the database servers in Lines 14--16.

\begin{algorithm}
\SetAlgorithmName{Algorithm}{}{}
\caption{ODES Init}\label{alg:odes_init}
\KwData{
A relation $R$ of cardinality $n$,
i.e., there are $|R| = n$ plaintext tuples;
a set of $m$ non-colluding nodes $N$;
function $Rnd()$ returning a random number;
}
\KwResult{
Each node $N_j$ holds a list of ciphertext shares $R_j$, $|R_j| = n$;
the index $idx$ holding the orders of $R$;
}
\nonl \; 
\For{i = 0; i < n; i++}{
    r = R[i] \;
    \For(\tcp*[f]{Share()}){j = 1; j < m; j++}{
        $s_i[j] \gets Rnd()$ \;
        $s_i[0] \coloneqq r - s_i[j]$ \;
    }
    Permute elements in $s_i$ \;
    \For{j = 0; j < m; j++}{
        Send $s_i[j]$ to $N_j$\;
        $R_j[i] \coloneqq s_i[j]$\tcp*{On server $N_j$}
    }
    Update $idx$\;
} 

\nonl \;
\For{j = 0; j < m; j++}{
    Send $idx$ to $N_j$\;
}
\end{algorithm}

The complexity of Alg.~\ref{alg:odes_init} is as follows.
Lines 3--6 take $\mathcal{O}(m)$ steps,
Line 7 takes $\mathcal{O}(m!)$ steps,
and Lines 8--11 take $\mathcal{O}(m)$ steps.
Therefore, Lines 1--13 take $\mathcal{O}(nm!)$.
Lines 14--16 trivially take $\mathcal{O}(m)$ steps and can be ignored.
The total complexity of Alg.~\ref{alg:odes_init} is therefore $\mathcal{O}(nm!)$.
Although the $\mathcal{O}(m)$ factor in the asymptotic complexity seems costly,
it is usually not an issue in practice because $m$ is taken at a relatively small value, e.g., 2, 4, and 8.

\subsubsection{Insertion}

The insertion protocol assumes that the cluster of database servers already holds secret shares and will take in a new record from the client.
In this context, the protocol comprises two phases:
(i) the client prepares the secret shares of the new record and sends them to the cluster of servers, and
(ii) the servers update their local shares and the index metadata for ordering information.
We depict both phases in Alg.~\ref{alg:odes_insert}.

\paragraph{Client protocol}
Similarly to the initialization protocol,
the client splits the given record $r$ into $m$ shares in Lines 1--4.
The client then permutes the shares and then sends each of them to a distinct database server (Lines 5--8).

\paragraph{Server protocol}
A server always inserts the received share into its local table,
as shown in Line 9.
Lines 10--25 work on updating the index through a binary search for the correct order of the new record.
Line 13 computes the difference between the received share and the share whose corresponding plaintext is the median.
Line 14 allows all database servers to learn about the differences so that they can decide whether to move to the smaller or the larger half of the sorted shares (Lines 16--19).
If there exists a duplicate value in the database,
then the protocol will end up at Line 21 ($\delta = 0$);
otherwise, Line 24 sets the order of the new record.

\begin{algorithm}
\SetAlgorithmName{Algorithm}{}{}
\caption{ODES Insert}\label{alg:odes_insert}
\KwData{
A new record $r$ of record identifier $rid$ to be inserted into a relation $R$, $|R| = n$;
$R$ is not directly accessible and can only be reconstructed from $R_j$, 
each of which is stored at node $N_j$, 
$1 \le j \le m$;
a global index $idx$ holding the order information of $R$;
}
\KwResult{
$R_j$ is updated with an additional record;
$Inx$ is updated to reflect the new order;
}

\nonl \;
\Comment{On client} 
\For(\tcp*[f]{Share()}){j = 1; j < m; j++}{
    $s[j] \gets Rnd()$ \;
    $s[0] \coloneqq r - s[j]$ \;
}
Permute elements in $s$ \;
\For{j = 0; j < m; j++}{
    Send $s[j]$ to $N_j$\;
}

\nonl \;
\Comment{On server $N_j$}
$R_j \coloneqq R_j \cup \{s[j]\}$ \tcp*{Update the data share}
$lo \coloneqq 0$, 
$hi \coloneqq n-1$,
$\delta \coloneq 0$\;
\While(\tcp*[f]{Binary search for order position of $r$}){$lo < hi$}{
    $mid \coloneqq \lfloor \frac{lo + hi}{2} \rfloor$\;
    $\delta_j \coloneqq s[j] - R_j[idx[mid]]$\;
    Broadcast $\delta_j$ to $N$'s \tcp*{Sharing $\delta$ is fine, but not $s[j]$}
    $\delta \coloneqq \sum_{j=0}^m \delta_j$\;
    \uIf(\tcp*[f]{Compare()}){$\delta > 0$}{
        $lo \coloneqq mid + 1$\;
    }
    \uElseIf(\tcp*[f]{Compare()}){$\delta < 0$}{
        $hi \coloneqq mid - 1$\;
    }
    \uElse{
        insert($idx$, $mid$, $rid$)\tcp*{Found duplicate values in $R$}
    }
}
\If{$\delta \not= 0$}{
    insert($idx$, $low$, $rid$) \tcp*{No existing values found in $R$} 
}
\end{algorithm}

\paragraph{Complexity analysis}
The complexity of Alg.~\ref{alg:odes_insert} is as follows.
Lines 1--4 take $\mathcal{O}(m)$ steps,
Line 5 takes $\mathcal{O}(m!)$ steps,
and Lines 6--8 take $\mathcal{O}(m)$ steps.
Therefore, the client takes overall $\mathcal{O}(m!)$ steps.
Lines 11--12 and 16--21 imply $\mathcal{O}(\log n)$ iterations.
Line 14 takes $\mathcal{O}(m)$ steps.
Lines 21 and 24 take constant steps.
Therefore, the server takes overall $\mathcal{O}(m\log n)$ steps.

\subsubsection{Query}

We represent the order-related predicate as $OP$,
which is part of the query.
There are three phases in the query protocol:
\begin{itemize}
    \item In the first phase of the query protocol,
    the client (i.e., the provider in Fig.~\ref{fig:arch}) simply broadcasts the predicate to all database servers.
    \item The second phase happens on each database server,
    which works on the local shares that qualify for the $OP$ predicate and send them back to the client.
    \item In the third phase, the client (i.e., the customer in Fig.~\ref{fig:arch}) collects all the qualified shares and reconstructs the plaintexts.
\end{itemize}
These three phases are summarized in Alg.~\ref{alg:odes_query}.
Lines 5--8 check every indexed order to see whether the local share qualifies for $OP$ and if so,
the server $N_j$ replies its local share $R_j[k]$ to the client.
Lines 13--19 reconstruct the list of plaintexts by aggregating the shares into each of the tuples. 

\begin{algorithm}
\SetAlgorithmName{Algorithm}{}{}
\caption{ODES Query}\label{alg:odes_query}
\KwData{
An order-predicate $OP(\cdot)$ based on which a set of records are returned from $R$, $|R| = n$;
each of share of $R_j$ is stored at node $N_j$, 
$1 \le j \le m$;
}
\KwResult{
The client receives a set of records $r$'s whose order $r_{ord}$ satisfying $OP(r_{ord}) = True$;
}

\nonl \;
\Comment{On client}
\For{j = 0; j < m; j++}{
    Send $OP(\cdot)$ to $R_j$\;
}

\nonl \;
\Comment{On server $N_j$}
Receive $OP(\cdot)$ from client\;
\For{i = 0; i < n; i++}{
    \If{$OP(i)$}{
        $k \coloneq idx[i]$\;
        Send $R_j[k]$ to the client \;
    }
}

\nonl \;
\Comment{On client}
Collect $K$ values from each of $m$ nodes\;
res = []\;
\For{k = 0; k < K; k++}{
    $val \coloneq 0$\;
    \For(\tcp*[f]{Reconstruct()}){j = 0; j < m; j++}{
        $val$ += $R_j[k]$\;
    }
    res.append($val$)\;
}
return res\;

\end{algorithm}

The complexity of Alg.~\ref{alg:odes_query} is as follows.
For the server, the overall complexity is $\mathcal{O}(n)$ (Lines 5--10).
For the client, 
Lines 1--3 take $\mathcal{O}(m)$ steps.
Line 11 itself takes $\mathcal{O}(nm)$ steps because $K \le n$.
Lines 13--19 take $\mathcal{O}(nm)$ also.
Therefore, the overall complexity of the client is $\mathcal{O}(nm)$.

\subsubsection{Deletion and Modification}

The deletion of ODES is more straightforward than insertion.
The first phase is asking every database server to remove the corresponding local shares.
The second phase is to remove the order information of the to-be-removed record in the index file and update the index file by decrementing (by one) the order values of those records whose orders are larger than the order of the deleted record.
We omit the detailed description of this protocol in this conference paper.

The modification of ODES can be trivially implemented by first deleting the record and then inserting the new value. 
We skip the formal protocol description of the modification in this paper.

\subsection{Security Analysis}\label{sec:security}

\subsubsection{Threat Model}

The database servers are assumed to be semi-honest,
meaning that they are not interested in launching active attacks such as compromising the secret shares.
However, those servers may eavesdrop on the messages and share passively;
for example, the servers may manage to collect secret shares of arbitrary plaintexts,
which essentially implies a chosen-plaintext attack (CPA).

In the context of order-preserving data confidentiality, however,
achieving the conventional CPA indistinguishability (IND-CPA) is impossible because the ordering information does allow the adversary to make a guess much better than a random one.
Therefore, we assume the adversary server is interested in learning information other than the orders of the plaintext.
This is also called IND-OCPA in the literature.
A stronger security notion was proposed called indistinguishability under frequency-analyzing ordered chosen-plaintext attack (IND-FAOCPA)~\cite{fker_ccs15}, 
implying that the access pattern cannot be leaked.

\subsubsection{Security Assumption}

We assume that a block cipher (e.g., AES) can be used as a pseudorandom generator (PRG) in practice.
We need this assumption because random numbers (i.e., the function $Rnd()$) are essential for our protocols (see Algs.~\ref{alg:odes_init}--\ref{alg:odes_query}).
This is a well-accepted assumption in the literature of applied cryptography.

\subsubsection{Indistinguishability between Two Parties}

Let $m_L$ and $m_R$ denote two distinct plaintext messages of the same length $l$.
According to Alg.~\ref{alg:odes_init},
when $m_L$ is encrypted, 
$s_0 \gets Rnd()$ and $s_1 \coloneq m_L - s_0$.
The function can be similarly defined for $m_R$.
Because we assume two nodes $N_0$ and $N_1$ do not collude,
only one share can be accessed.
Formally, we define the following function
\[\displaystyle
ODES_L^2(m_L, m_R, i\in \{0, 1\}) \defeq 
\begin{cases}
s_1 = Rnd(), & i = 1 \\
s_0 = m_L - s_1, & i = 0
\end{cases}
\]
and
\[\displaystyle
ODES_R^2(m_L, m_R, i\in \{0, 1\}) \defeq 
\begin{cases}
s_1 = Rnd(), & i = 1 \\
s_0 = m_R - s_1. & i = 0
\end{cases}
\]

The goal is to show that when an adversary $\mathcal{A}$ submits two messages $m_L$ and $m_R$ to the ODES oracle,
$\mathcal{A}$ cannot distinguish whether a share $s_i$ is from $ODES_L^2$ or $ODES_R^2$,
$i \in \{0, 1\}$.
If $s_1$ is accessed, 
then it is obvious that $s_1$ cannot be distinguished because both $ODES_L^2$ and $ODES_R^2$ return random numbers through $Rnd()$.
If $s_0$ is accessed,
we need to show that $m_L - Rnd()$ and $m_R - Rnd()$ are indistinguishable. 
This is indeed the case because $Rnd()$ is supposed to be uniformly distributed in the message space and adding its output to $m_L$ or $m_R$ would render it garbled.
Technically, the probability for $\mathcal{A}$ to distinguish two randomized numbers is $\frac{1}{2^l}$,
which is a negligible function in $l$---the bit length of the plaintext.
In other words, the $s_0$ shares generated by $ODES^2_L$ and $ODES^2_R$ are interchangeable.

\subsubsection{Indistinguishability between Arbitrary Parties}

In the general case, $t$ nodes split the plaintexts,
the secrets can be similarly defined in extended functions:
\[\displaystyle
ODES_L^t(m_L, m_R, i\in [0, t)) \defeq 
\begin{cases}
s_i = Rnd(), & i \not= 0 \\
s_0 = m_L - \sum_{i=1}^{t-1} s_i, & i = 0
\end{cases}
\]
and
\[\displaystyle
ODES_R^t(m_L, m_R, i\in [0, 1)) \defeq 
\begin{cases}
s_i = Rnd(), & i \not= 0 \\
s_0 = m_R - \sum_{i=1}^{t-1} s_i. & i = 0
\end{cases}
\]
When $i \not= 0$, the $s_i = Rnd()$ values cannot be distinguished between $ODES^t_L$ and $ODES^t_R$.
When $i = 0$, repeated $Rnd()$'s can only further garble the $m_L$ and $m_R$ messages (recall that the arithmetic operation over a set of negligible functions results in a negligible function as well),
making $s_0$ interchangeable.

\subsubsection{IND-FAOCPA}

To show that ODES is IND-FAOCPA,
we need to demonstrate two features:
(i) releasing the access patterns does not help the adversary to distinguish between two ciphertexts, and 
(ii) the sequence of pairs of plaintext messages is always ordered such that the adversary has no extra information to help the distinguishing process.
We will show that both properties are satisfied by ODES.
A complete proof is beyond the scope of this conference paper and we will keep our analysis at the descriptive level.

\paragraph{Frequency Analyzing}
Every time a plaintext interacts with the cluster of database servers,
the plaintext is freshly decomposed by the \textit{Share()} primitive (\S\ref{sec:primitive}).
The probability that the shared secrets are repeated is negligible, 
$\frac{1}{2^{l\cdot m}}$,
where $l$ denotes the bit-string length of the message space and $m$ denotes the number of database servers.
Intuitively, this can be understood as the chance of generating a repeated set of shares given the same plaintext is extremely low. 
Therefore, the adversary cannot launch a successful attack by studying the frequency or access patterns.

\paragraph{Ordered Plaintexts}
Recall that one assumption of the proposed ODES scheme is that they are non-colluding. 
This implies that the database servers have no way to learn about the plaintext except for the ordinal information of its local shares.
However, as we discussed in~\S\ref{sec:security},
those shares are indistinguishable from a random number in the local table of the database server.
For example, in Figure~\ref{fig:odes_example},
although the positive database server knows that 22-MAR has a higher balance than that of 22-FEB (by collecting the $\delta$s), 
there is no way for the server to reveal the real balance of either 22-MAR or 22-FEB (unless both servers send over the local share in its entirety, which is not allowed in ODES).
The order of local shares on an individual database server does not help;
for example, even the servers themselves cannot reach a consensus (let alone whether the order is consistent with the plaintext):
the positive server says the local share of 22-MAR (\$11,000) is lower than that of 22-FEB (\$14,000) and yet the negative server says the local share of 22-MAR (\$2,000) is higher than that of 22-FEB (\$-2,000).

\section{Evaluation}

\subsection{System Implementation}

We have implemented the proposed ODES protocol with about 1,200 lines of Python code and Bash script,
which will be released at
\url{https://github.com/hpdic/odes}.
We choose the lightweight SQLite as the local database instance.
Note that SQLite is a file-based database and does not support network access.
We thus implement a communication layer among remote SQLite instances through the \textit{paramiko} library for secure data transfer and remote query invocation.
Some of the most important libraries and dependencies include:
\textit{python} 3.8.0,
\textit{sqlite} 3.31.1,
\textit{numpy} 2.21.0,
\textit{paramiko} 2.12.0,
\textit{scp} 0.14.4, 
and \textit{cryptography} 39.9.0.

\subsection{Experimental Setup}

\subsubsection{Test Bed}

We deploy the proposed ODES protocol and the latest OPE scheme~\cite{dli_vldb21} with SQLite~\cite{sqlite} on a 10-node cluster hosted at CloudLab~\cite{cloudlab}.
Each node is equipped with two 32-core Intel Xeon Gold 6142 CPUs, 384 GB ECC DDR4-2666 memory, and 	
two 1~TB SSDs.
The operating system image is Ubuntu 20.04.3 LTS,
and the page size is 4 KB.
All servers are connected via a 1 Gbps control link (Dell D3048 switches) and a 10 Gbps experimental link (Dell S5048 switches).
We only use the experimental links for our evaluation.

Specifically, we name the 10 nodes in the cluster as \textit{node0}--\textit{node9}.
The client runs on \textit{node0},
the ODES server runs on \textit{node1}--\textit{node8},
and the OPE server runs on \textit{node9}.
All 10 nodes are enabled with password-less SSH connection for convenient communication since our evaluation focuses on performance metrics rather than security measurement. 
Figure~\ref{fig:client_overhead} illustrates the topological structure of our 10-node cluster.

\begin{figure}[!t]
  \centering
  \includegraphics[width=\linewidth]{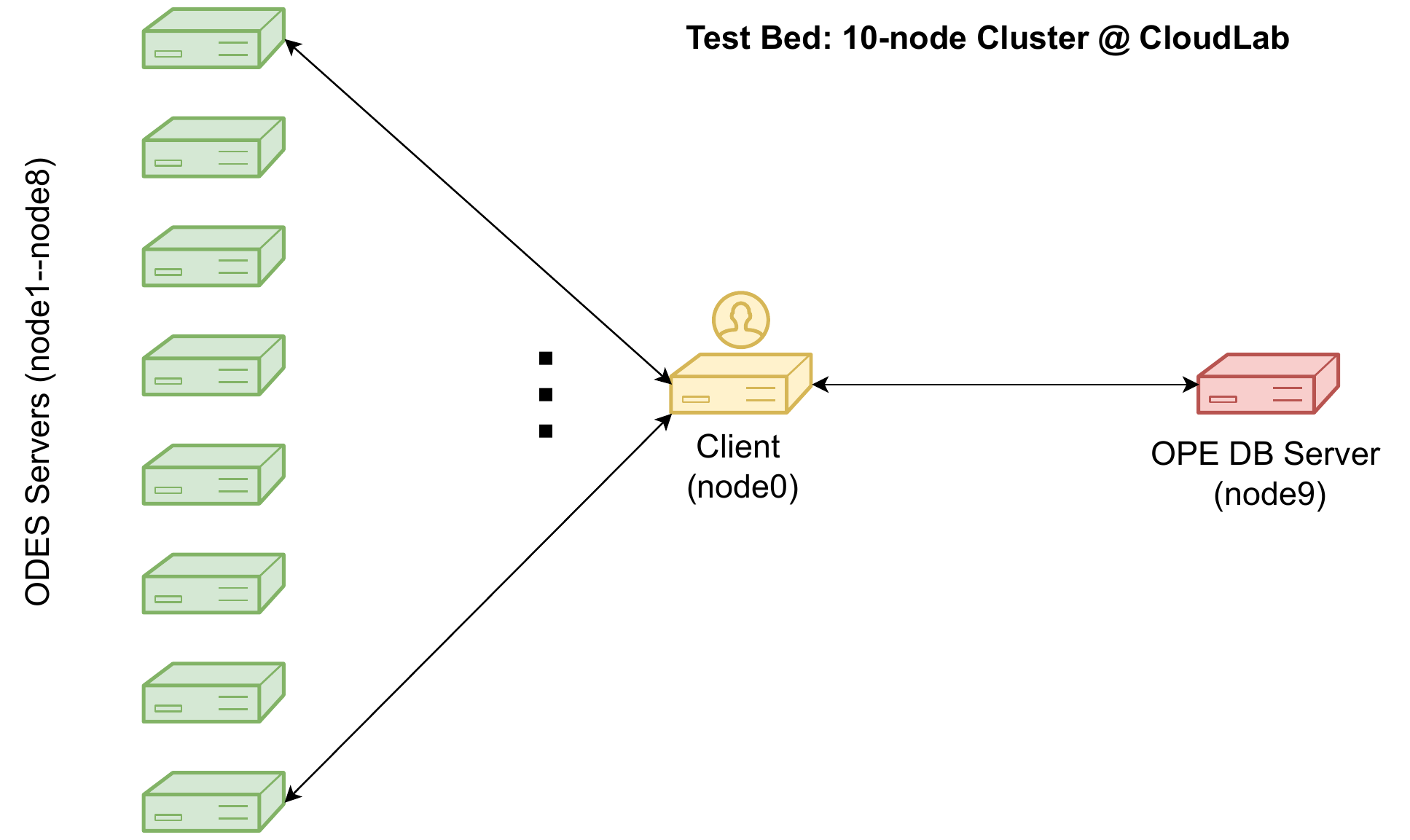}
  \caption{10-node cluster on CloudLab}
  \label{fig:cloudlab}
\end{figure}

\subsubsection{Baseline Systems}

The literature on order-preserving encryption (OPE) is mainly contributed by the security and database communities.
Not all proposed ideas had been implemented;
as of the writing of this paper,
the state-of-the-art OPE system was presented at~\cite{dli_vldb21},
which adopts the symmetric encryption scheme 128-bit AES~\cite{aes} as a building block.
Because the plaintext is not always 128-bit long,
our implementation takes the PKCS7~\cite{pkcs7} padding scheme.
We will use the term OPEA to refer to this particular system.

The original OPEA system was implemented with a server user-defined function on MySQL and a client script using Python.
However, to make a fair comparison, we re-implement both the server and client functions of OPEA for SQLite;
in particular, we leverage the state-of-the-art secure communication framework \textit{paramiko} and the cryptographic package \textit{hazmat} for security building blocks in our implementation.

\subsubsection{Data Sets}
\label{sec:eval_dataset}

The synthetic benchmark is TPC-H ver. 3.0.0~\cite{tpch3}, a standard database benchmark.
There are overall eight tables in TPC-H;
we select four of them for evaluating the proposed work:
Supplier, Customer, Part, and Orders.
The reason why we choose these four tables is two-fold.
First, they all exhibit a single-attribute primary key that is straightforward to encode in the underlying data structures.
Second, they all have at least one numerical attribute that is more interesting to encode than textual attributes.
The attributes we are interested in include:
\begin{verbatim}
    Supplier.S_Acctbal, 
    Customer.C_Acctbal, 
    Part.P_Retailprice, 
    Orders.O_Totalprice.    
\end{verbatim}
In our TPC-H of scale 0.01, the above column includes 100, 1,500, 2,000, and 15,000 tuples, respectively.

In addition to the synthetic benchmark, 
we also evaluate the proposed work with three real-world data sets.
\begin{itemize}
    \item COVID-19. The first application is the U.S. national COVID-19 statistics from April 2020 to March 2021~\cite{covid19data}.
    The data set has 341 days of 16 metrics, such as \textit{death increase}, \textit{positive increase}, and \textit{hospitalized increase}.
        
    \item Bitcoin. The second application is the history of Bitcoin trade volume~\cite{bitcoin_trade} since it was first exchanged in the public in February 2013.
    The data consists of the accumulated Bitcoin exchange on a 3-day basis from February 2013 to January 2022,
    totaling 1,086 large numbers.
    
    \item Human Genome \#38 (hg38). The third application is the human genome reference 38~\cite{hg_data}, 
    commonly known as \textit{hg38},
    which includes 34,424 rows of singular attributes,
    e.g., \textit{transcription positions}, \textit{coding regions}, and \textit{number of exons}, last updated in March 2020.
\end{itemize}

\subsubsection{Workloads}

As discussed in~\cite{dli_vldb21},
we are primarily interested in two performance metrics in an order-preserving database encryption scheme:
insertion and query.
\begin{itemize}
    \item The insertion workload works as follows. 
    The client roughly issues the following SQL to insert a series of records into the SQLite database:
    \begin{verbatim}
INSERT INTO <table name> VALUES <ciphertext>;
    \end{verbatim}
    The insertion starts from scratch,
    i.e., the target table is empty.
    During the insertion, some sort of indexing (depending on the scheme, e.g., ODES, OPEA) is carried out for future queries touching on ordering/sorting operations.
    Essentially, the insertion incurs not only the searching overhead of comparing the provided ciphertext and those existing (encrypted) tuples already in the table but also the updating overhead for maintaining the order information after the new record is inserted.
    
    \item The query workload is simpler than the insertion. 
    We randomly make $\log(n)$ point-wise queries and compare their values,
    where $n$ denotes the total number of records in the table, 
    i.e., cardinality.
    Intuitively, if there are a large number of records for insertion,
    the query time should be significantly smaller than the insertion time.
\end{itemize}

We will also report the overhead incurred by the client, if applicable.
The client overhead includes the initialization and maintenance of encoding the plaintext and optionally other auxiliary information such as \textit{local table} in OPEA or \textit{linear secret sharing} in ODES. 

We carry out all experiments at least three times and report the average numbers.
We do not plot the variances (i.e., the error bars) because they are negligible.
A complete log of the experiments is available upon request.
Unless otherwise stated, the default number of ODES shares is two.

\subsection{End-to-end Performance on TPC-H}

\begin{figure}[!t]
  \centering
  \includegraphics[width=\bigfigwidth]{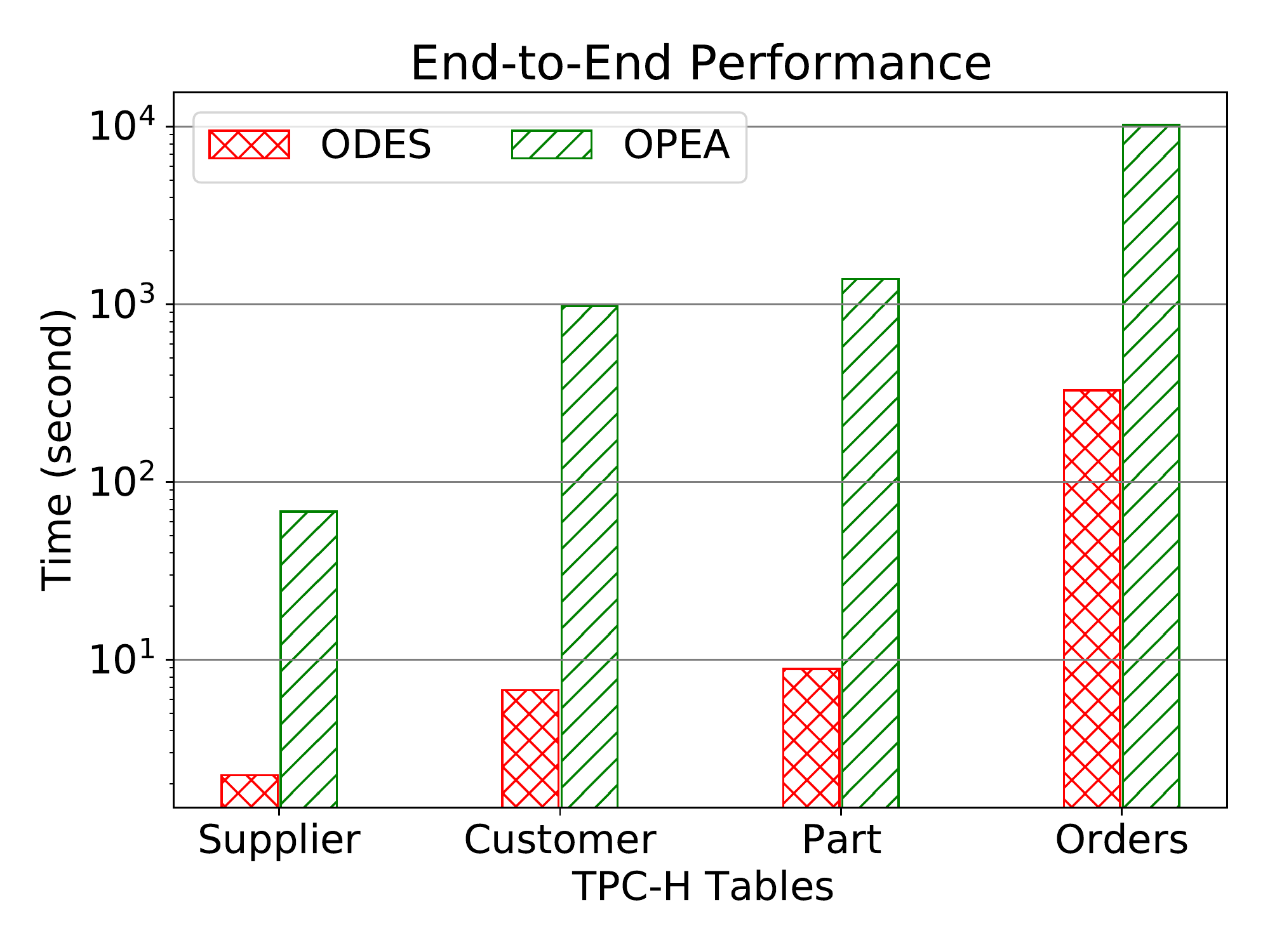}
  \caption{Overall performance of ODES and OPEA}
  \label{fig:overall}
\end{figure}

We start with reporting the end-to-end performance of the proposed ODES protocol with the conventional OPEA.
Figure~\ref{fig:overall} shows the end-to-end execution time of inserting $n$ tuples into an empty table followed by $\log(n)$ queries,
where $n$ denotes the total number of records as discussed in $\S$\ref{sec:eval_dataset}.
Four TPC-H tables are involved and the time includes the overhead on the client side in addition to the insertion and the query time.

According to the results of Figure~\ref{fig:overall},
ODES outperforms OPEA by orders of magnitude in all TPC-H tables.
Specifically, for Customer and Part tables,
the improvement is about 100$\times$;
for Supplier and Orders tables,
the speedup is over 10$\times$.
Therefore, we claim that ODES can improve the overall order-preserving encryption time $10\times$--$100\times$ faster on TPC-H.
In the following three subsections,
we will break down the overall running time into three phases:
client overhead, insertion time, and query time.

\subsection{ODES for Real-World Applications}

We report the ODES performance on three different real-world data sets.
We break down the execution time into client overhead, insertion time, and query time to have a better idea of the distribution of cost in real-world ODES applications.
Since our goal is to gain more insight into cost allocation in practice,
we will not compare the metric to the baseline, i.e., OPEA.

\begin{figure*}[!t]
     \centering
     \begin{subfigure}[b]{\subfigwidthThree\textwidth}
         \centering
         \includegraphics[width=\textwidth]{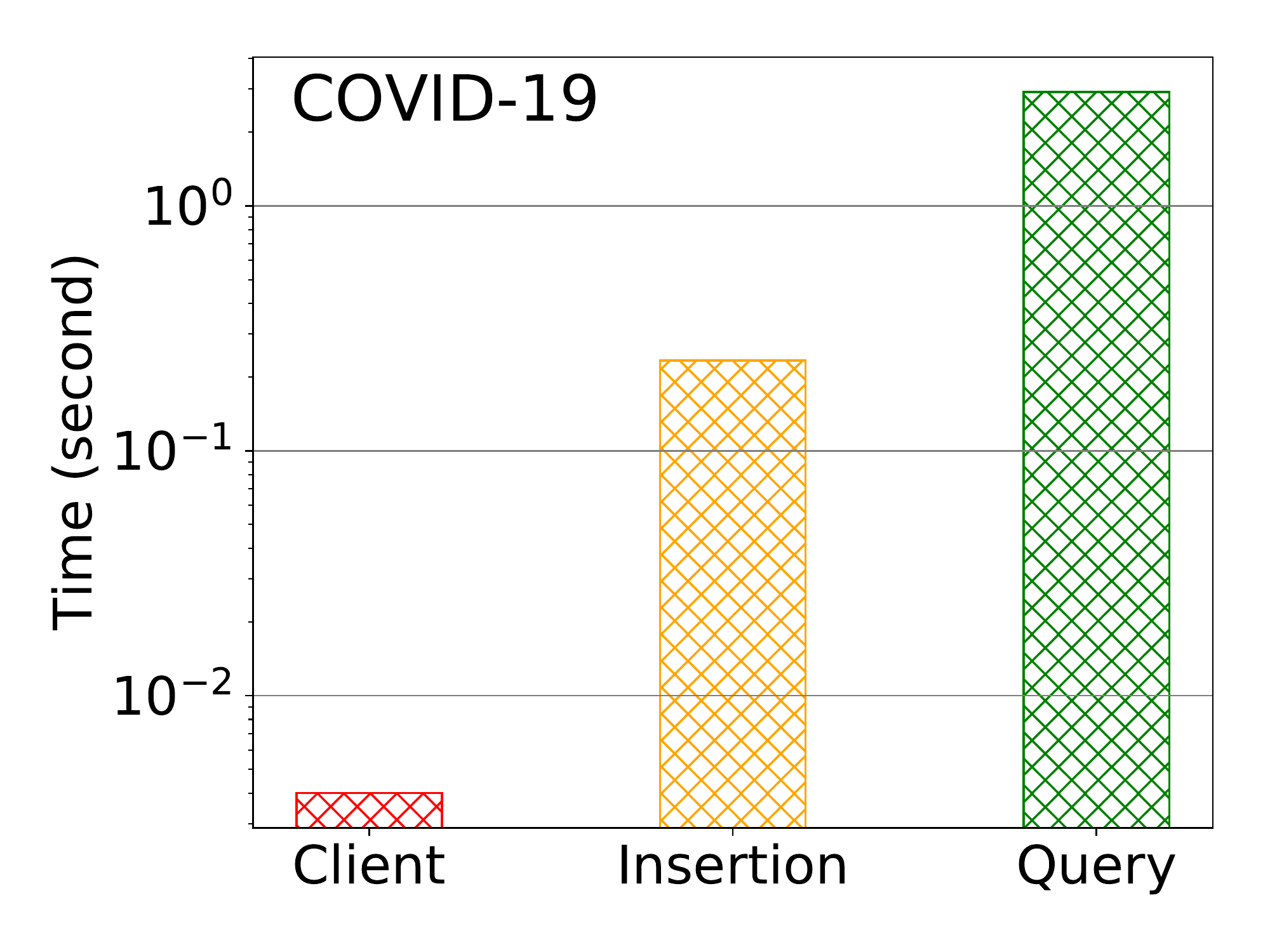}
     \end{subfigure}
     \hfill
     \begin{subfigure}[b]{\subfigwidthThree\textwidth}
         \centering
         \includegraphics[width=\textwidth]{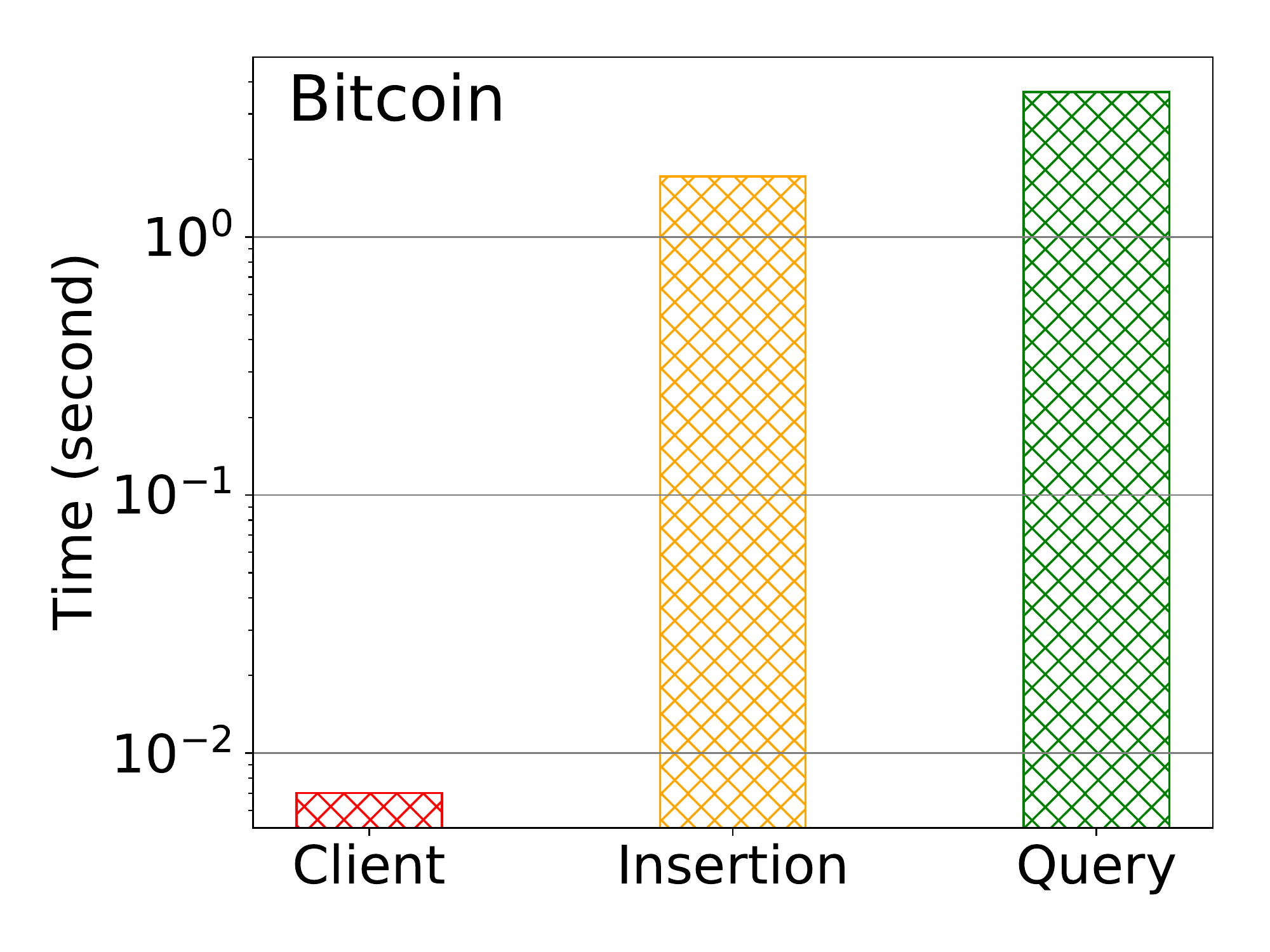}
     \end{subfigure}
     \hfill
     \begin{subfigure}[b]{\subfigwidthThree\textwidth}
         \centering
         \includegraphics[width=\textwidth]{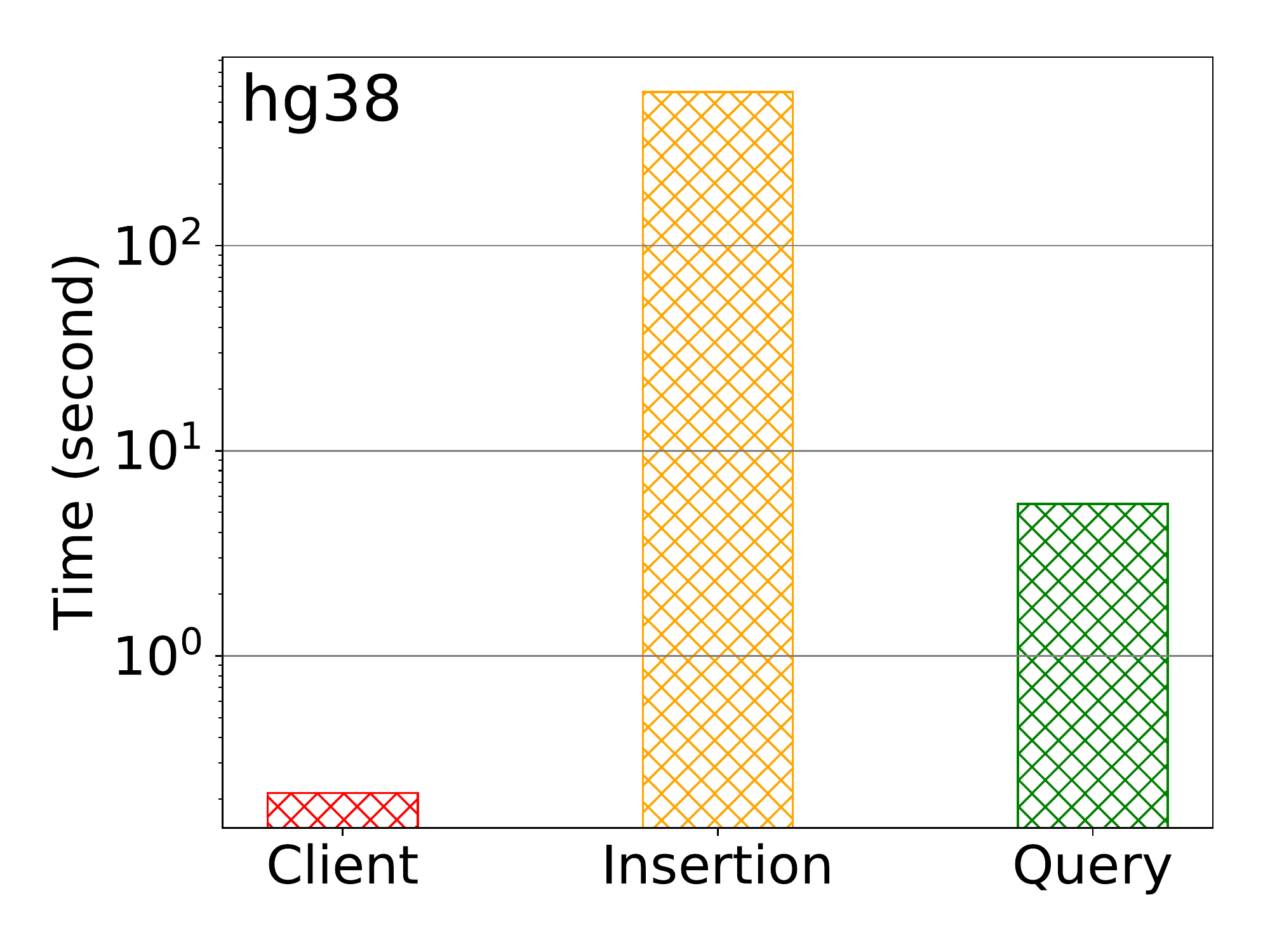}
     \end{subfigure}
        \caption{ODES for real-world applications}
        \label{fig:odes}
\end{figure*}

Figure~\ref{fig:odes} shows the time breakdown of those three real-world data sets.
There are a few interesting observations.
First, the client overhead is insignificant, 
if not negligible.
This reaffirms our previous conclusion on the effectiveness of ODES for lightweight devices in the Internet of Things (IoT) or edge computing applications.
Second, the cost allocation between insertion and query is dynamic.
For COVID-19, the insertion cost is almost 10$\times$ lower than the query cost;
however, a converse phenomenon occurs for hg38,
where the insertion cost is more than 10$\times$ higher than the query.
This can be best explained by the cardinality of the table:
there are 341 rows in COVID-19 and more than 30,000 records in hg38.

\subsection{Client Overhead}

Recall that in OPEA, the client maintains a key-value store of plaintext and ciphertext such that when inserting a new record the client can locate the ciphertexts of the lower and upper encoding.
This is required because OPEA must synchronize the stateful client-side table and the recording encoding on the server side.
Each ciphertext is essentially a 128-bit byte-string (because OPEA uses 128-bit AES for encryption) regardless of the plaintext length\footnote{Obviously, we here assume the tuple is represented by less than 128 bits. This is indeed the case because,
practically speaking,
the numerical values in database applications are almost always less than $2^{128}$.}.
Assuming a plaintext numerical value is stored with 4 bytes (32 bits),
the expansion rate of the local table in OPEA is
$\frac{128 + 32}{32} = 5\times$.

By contrast, the ODES client is stateless and lightweight:
(i) the client simply generates $m-1$ random numbers as shares and calculate the the $m$-th share by subtracting the plaintext with those $m-1$ random numbers;
(ii) the client does not maintain any data structure to store the intermediate values (technically, the expansion rate is zero if we follow the same terminology as OPEA).
Therefore, it is reasonable to expect a much lower client overhead in ODES.

\begin{figure}[!t]
  \centering
  \includegraphics[width=\bigfigwidth]{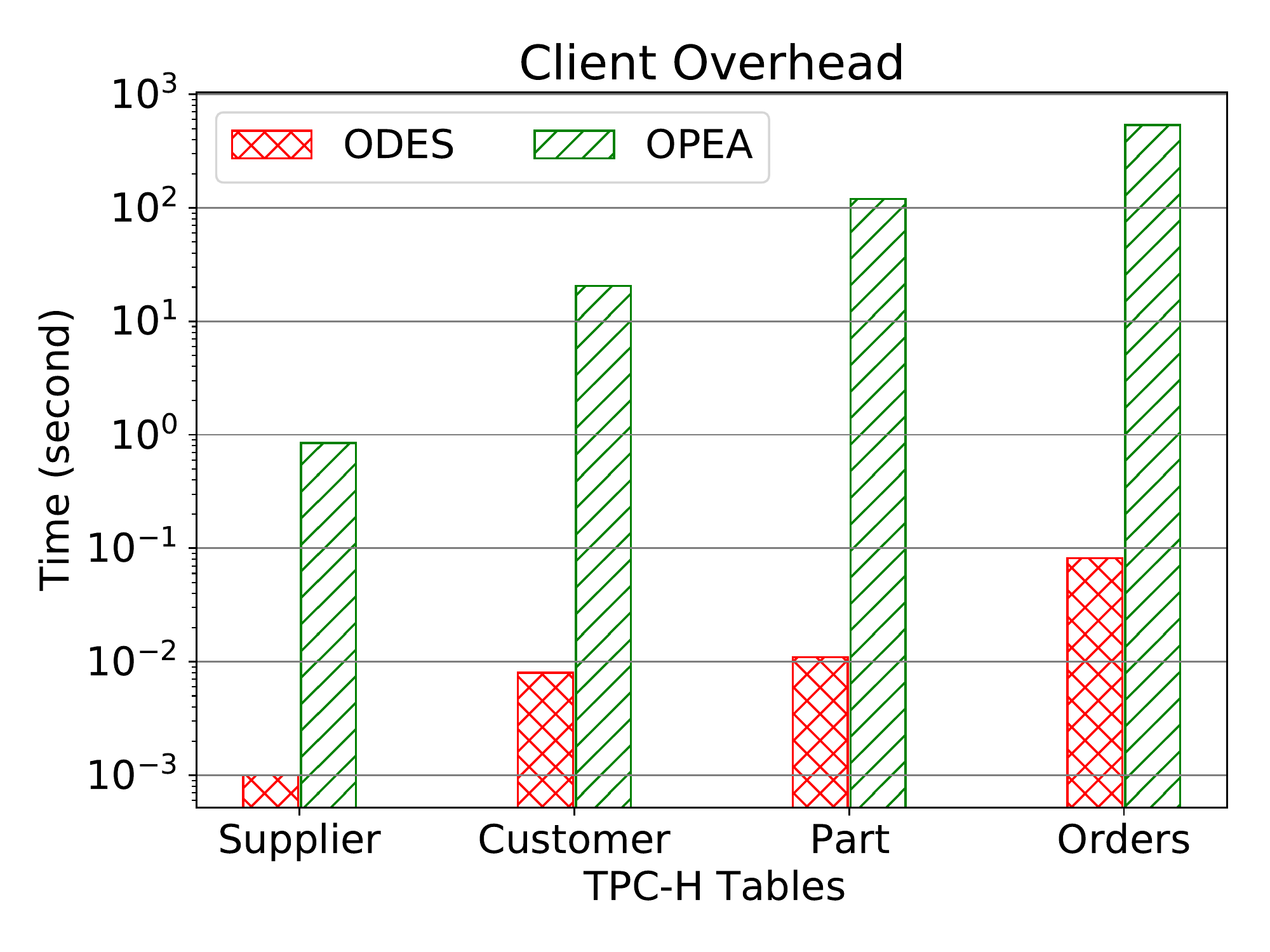}
  \caption{Client overhead of ODES and OPEA}
  \label{fig:client_overhead}
\end{figure}

Figure~\ref{fig:client_overhead} reports the overhead of ODES and OPEA on the client side.
Indeed, we observe that the ODES client overhead is about 3--4 orders of magnitude lower than OPEA for all TPC-H tables.
In addition to the lower overhead,
the results also suggest that ODES is a more practical solution for those lightweight devices such as smartphones and smartwatches,
which have limited computational power and storage capacity.

\subsection{Insertion Performance}

The main cost of inserting a record in OPEA lies in the interaction between the client and the server.
After the client locates the lower and upper ciphertexts that are closest to the next plaintext,
the client needs to make two queries to the server to retrieve the encoding of both ciphertexts.
For simplicity, the encoding can be thought of as the positions of the plaintexts.
Roughly speaking, one OPEA insertion incurs two additional database queries (one for the lower neighbor and the other one for the upper neighbor).
That is, the number of queries is tripled at the time of insertions.
In addition, the client needs to adjust the local table to incorporate the newly inserted plaintext and its associated ciphertexts, 
the entry position in the local table,
and the counter of possible repetition of duplicate plaintexts.

In contrast, the main cost of ODES for inserting records lies in the cost of uploading secret shares to multiple servers.
However, ODES is a stateless protocol,
meaning that no extra information needs to be maintained on either the client or the server side.
That is, the queries among distinct servers can be parallelized.
As a result, although the overall number of queries in ODES may exceed that in OPEA (if there are more than three ODES servers),
the running time for executing these queries is about the same as a single query.
Combining all the above factors,
we expect that ODES runs must faster than OPEA for inserting records into the database.

\begin{figure}[!t]
  \centering
  \includegraphics[width=\bigfigwidth]{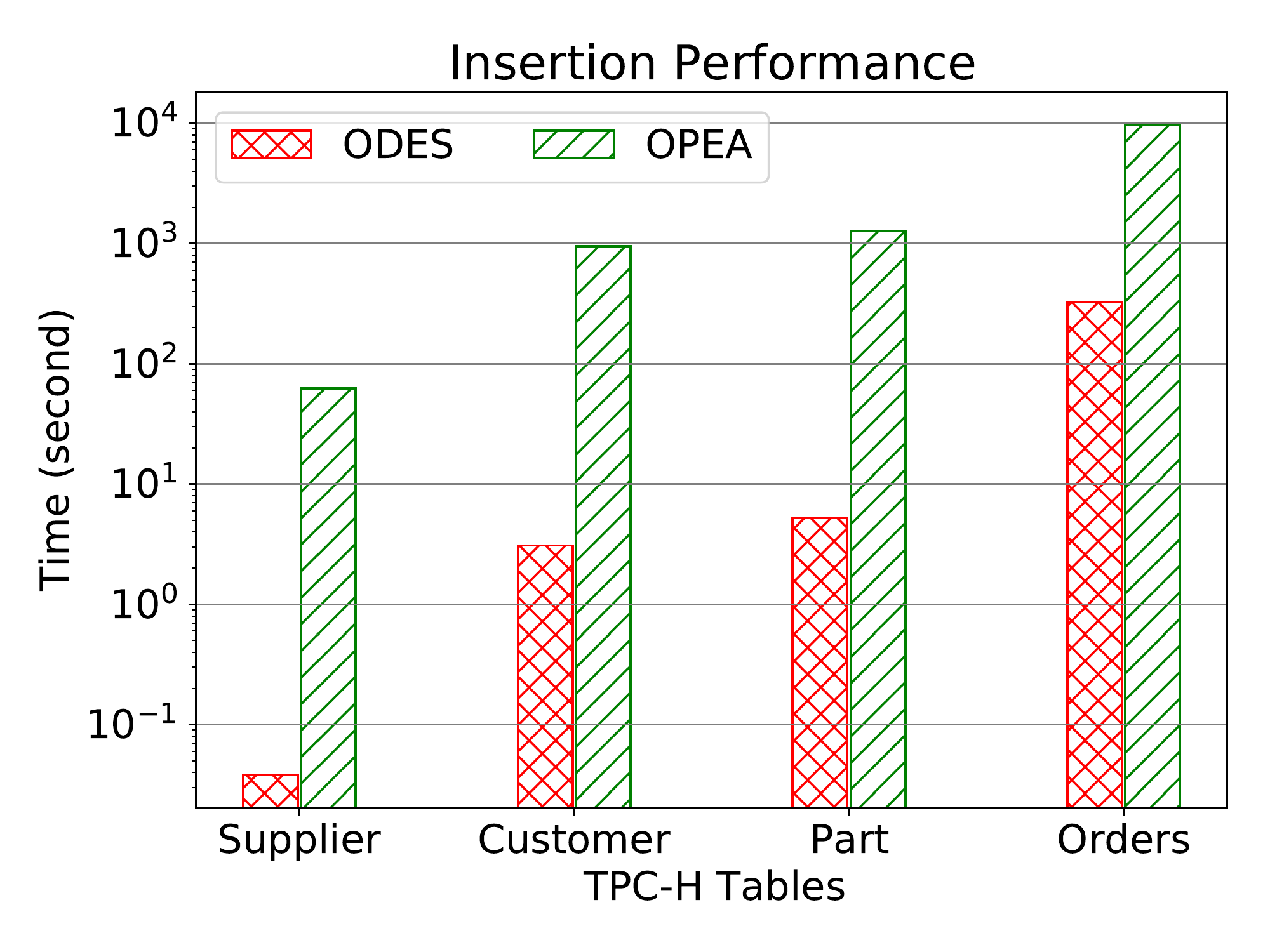}
  \caption{Insertion performance of ODES and OPEA}
  \label{fig:insert}
\end{figure}

Figure~\ref{fig:query} reports the insertion performance of ODES and OPEA on TPC-H tables.
We observe that for small and medium scales (Supplier, Customer, Part),
ODES outperforms OPEA by more than two orders of magnitude.
For the large-scale data set of the Orders table,
the speedup is also significant, 
exceeding 10$\times$.

\subsection{Query Performance}

The performance gap between ODES and OPEA for queries is significantly smaller than that for record insertion because the ordering information of OPEA is stored as plaintext and therefore an efficient binary sorting can be leveraged.
However, as discussed above,
the expanded ciphertext (i.e., 5$\times$) takes significantly more storage space and network transmission than the plaintext.
On the other hand,
ODES deals with $m\times$ plaintexts during the queries that can be parallelized,
where $m$ denotes the number of nodes (shares).

\begin{figure}[!t]
  \centering
  \includegraphics[width=\bigfigwidth]{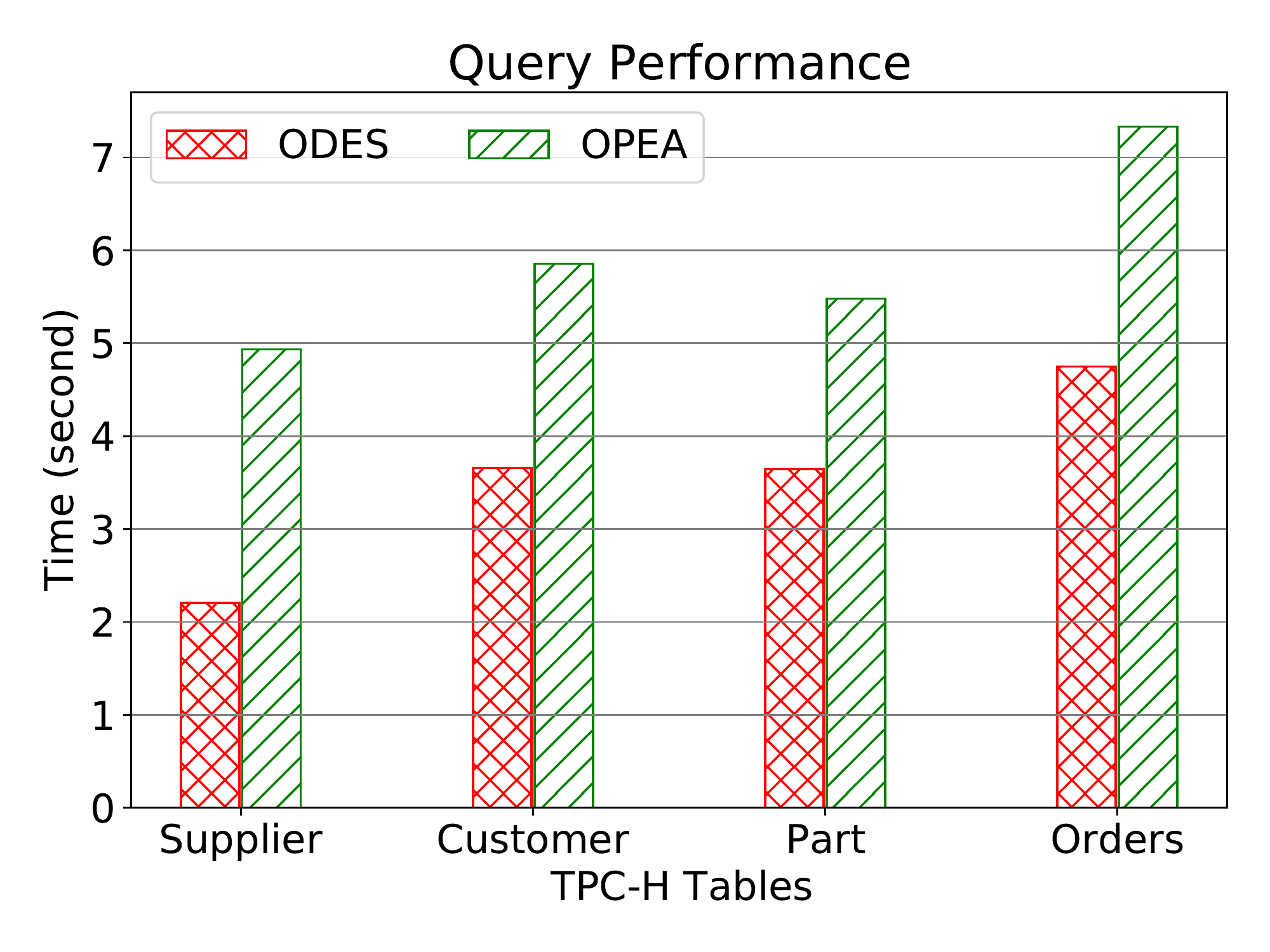}
  \caption{Query Performance of ODES and OPEA}
  \label{fig:query}
\end{figure}

Figure~\ref{fig:query} reports the query performance of ODES and OPEA.
We observe that ODES still outperforms OPEA in terms of query performance,
and yet the gap is only about 35\%--55\% rather than orders of magnitude as we have seen for insertion.
However, we argue that 35\%--55\% is nonetheless a significant improvement in query performance,
demonstrating the superiority of the proposed ODES protocol.

\subsection{Scalability of ODES}

The last metric we are interested in learning about ODES performance is the scalability regarding different numbers of shares.
Thus far, all experiments assume that there are two nodes,
each of which holds a secret share from the plaintext.
There is no technical limitation preventing the user from employing more nodes to increase the security level,
possibly by allowing more processing time.
That is, the adversary would need to compromise more nodes to launch a successful attack.

\begin{figure}[!t]
     \centering
     \begin{subfigure}[b]{\subfigwidth\textwidth}
         \centering
         \includegraphics[width=\textwidth]{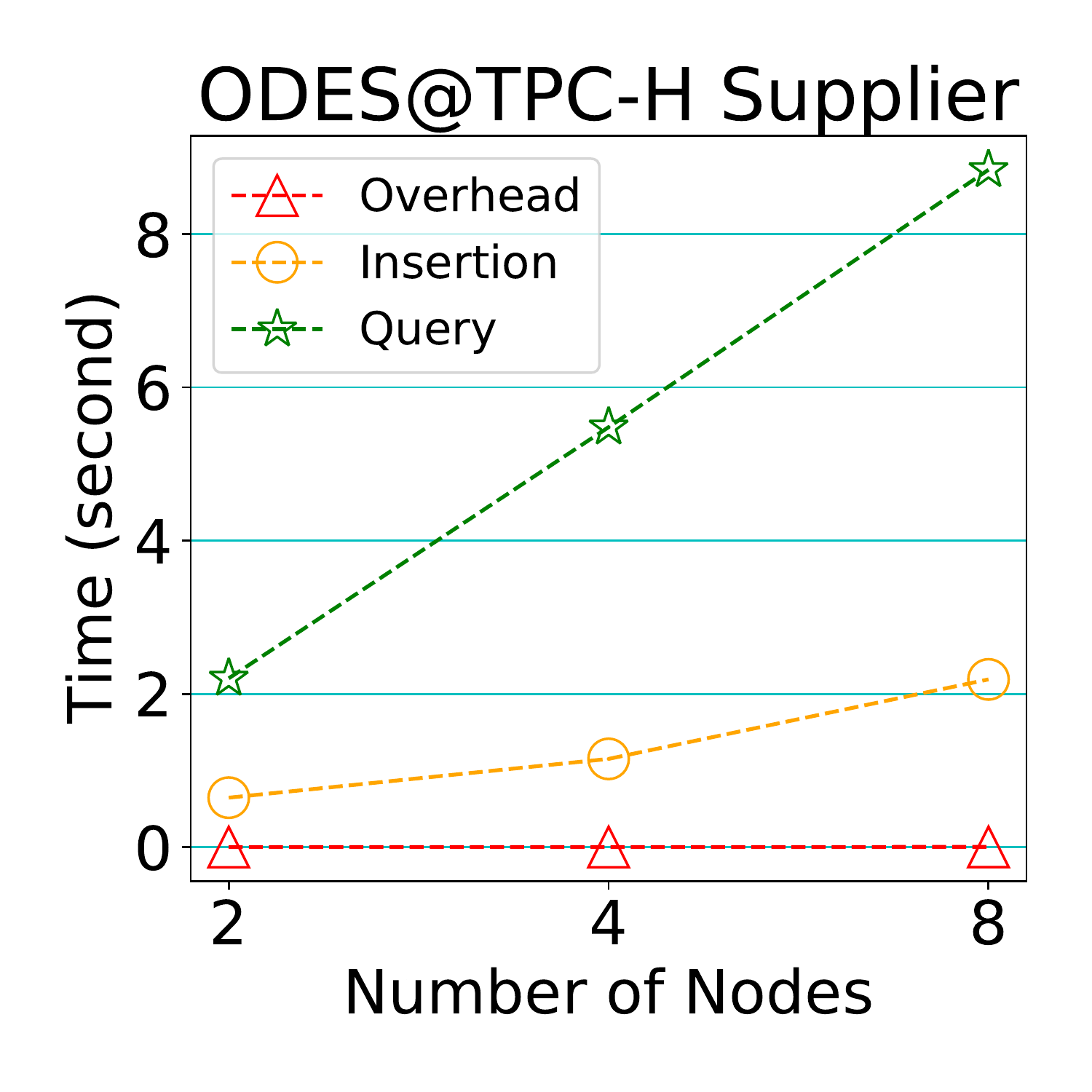}
     \end{subfigure}
     \hfill
     \begin{subfigure}[b]{\subfigwidth\textwidth}
         \centering
         \includegraphics[width=\textwidth]{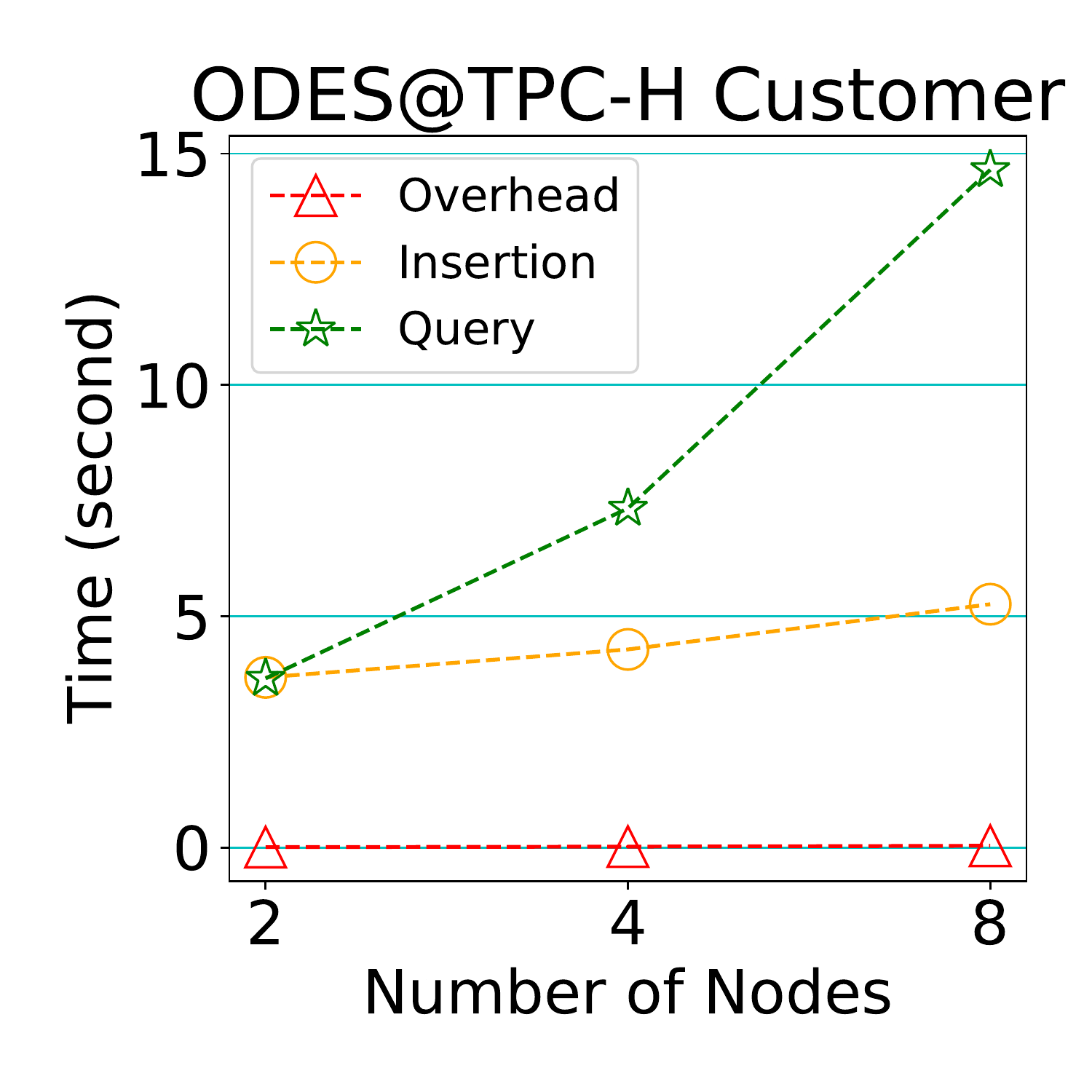}
     \end{subfigure}
    \\
     \begin{subfigure}[b]{\subfigwidth\textwidth}
         \centering
         \includegraphics[width=\textwidth]{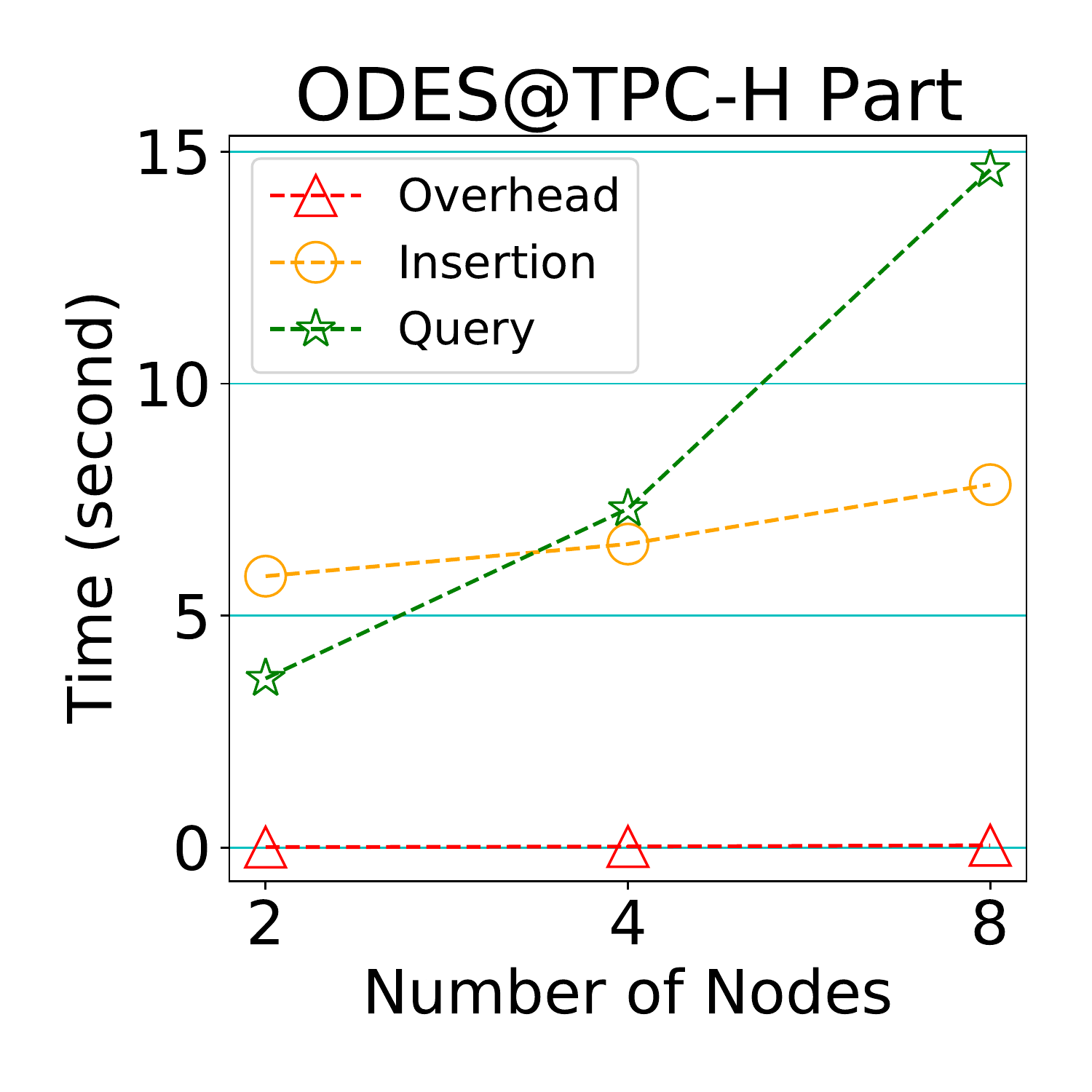}
     \end{subfigure}
     \hfill
     \begin{subfigure}[b]{\subfigwidth\textwidth}
         \centering
         \includegraphics[width=\textwidth]{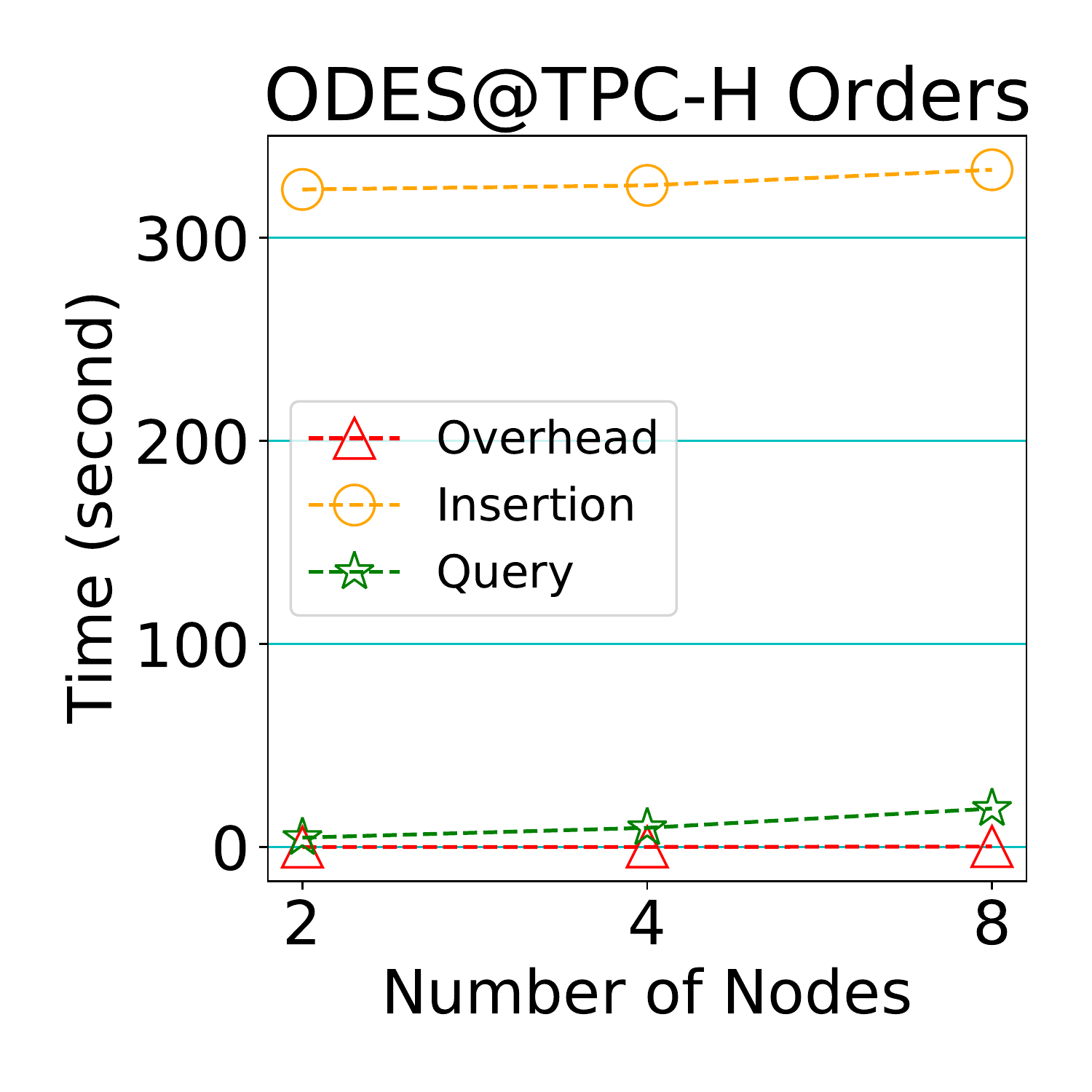}
     \end{subfigure}     
        \caption{Scalability of ODES on four TPC-H relations: Supplier, Customer, Part, and Orders.}
        \label{fig:scale}
\end{figure}

Figure~\ref{fig:scale} reports the scalability of ODES when various numbers of shares (i.e., nodes) are involved on four TPC-H tables.
We observe that the client overhead is negligible at all scales,
which is consistent with what we have learned.
The insertion cost mildly increases when more shares are involved,
which can be explained by the higher cost of splitting the plaintext into a larger number of shares.
The query time exhibits a steeper trend concerning the number of nodes;
the cost looks proportional to the number of nodes.
The main reason for this linear proportion is due to the serial implementation of the \textit{paramiko} library,
which is synchronous and therefore a remote query must complete before another one is issued.
We believe parallel processing from the client would greatly mitigate this issue by utilizing more CPU cores and higher network bandwidth,
which will be explored in our future work.

\subsection{Storage Cost}

\paragraph{Client Storage}
Recall that ODES is a stateless protocol,
meaning that there is zero storage requirement for ODES.
On the other hand, the space overhead of OPEA is $\mathcal{O}(\widehat{n})$,
where $\widehat{n}$ denotes the number of distinct values in the column.

\begin{figure}[!t]
  \centering
  \includegraphics[width=\bigfigwidth]{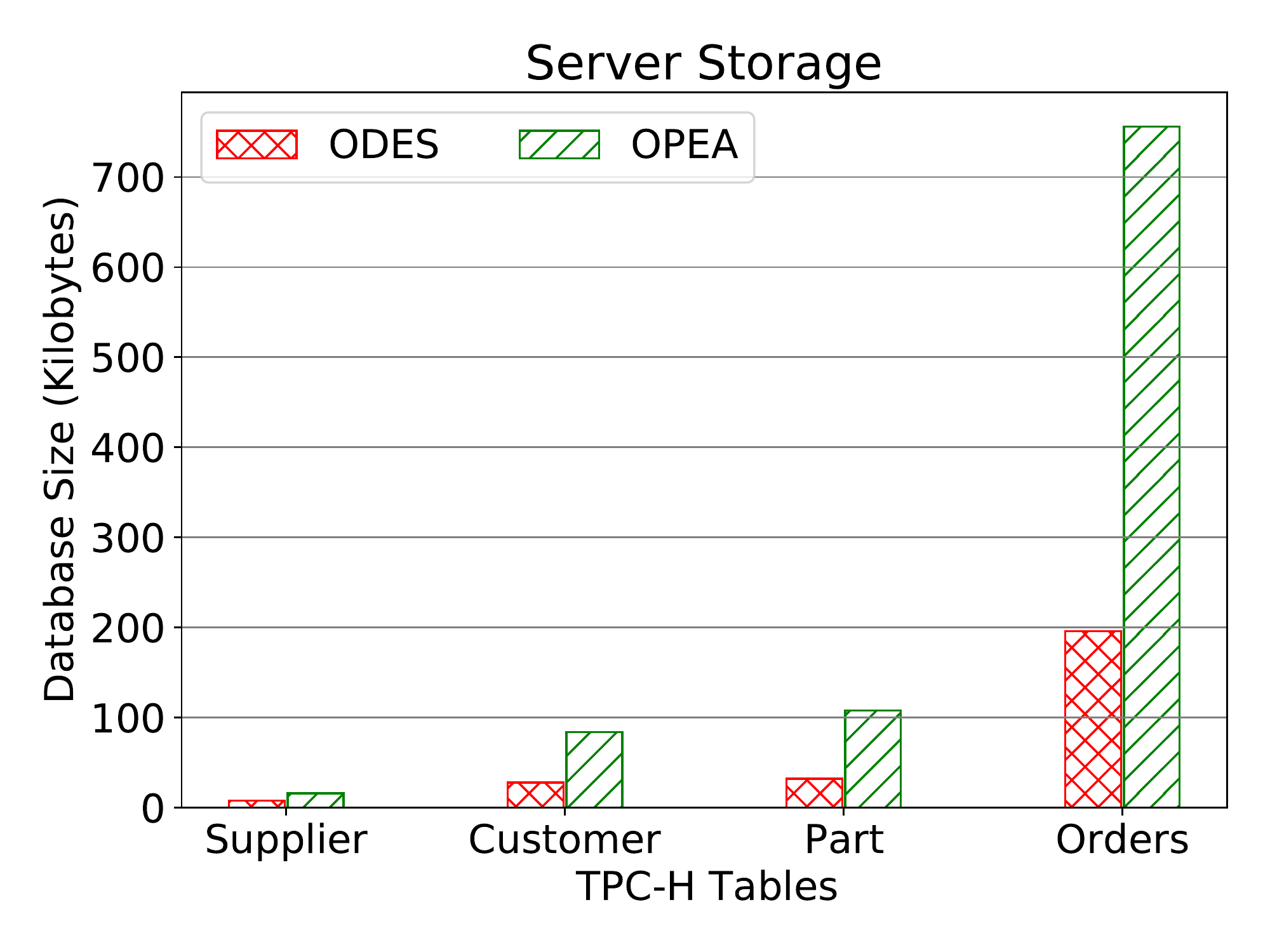}
  \caption{Server storage of ODES and OPEA}
  \label{fig:storage}
\end{figure}

\paragraph{Server Storage}
Figure~\ref{fig:storage} reports the database size of ODES and OPEA when various TPC-H tables are encrypted.
The results suggest that for small- and medium-scale data sets,
the database size is not significant.
However, for larger data sets such as the orders table,
the space overhead is much higher in OPEA.
This can be best explained by the fact that OPEA involves the AES scheme,
which causes size expansion in the ciphertext.
That is, no matter how small the plaintext value is,
the encrypted text will always be a fixed length (e.g., 128 bits) for security reasons.
Therefore, the more plaintext we have, the larger space overhead is incurred.

\section{Conclusion and Future Work}

This paper proposes a new stateless order-preserving encryption scheme,
namely ODES (Ordered Database Encryption with Secret-sharing),
by incorporating secret-sharing primitives.
ODES supports the latest IND-FAOCPA security.
A series of database protocols are designed based on ODES.
The ODES scheme and database protocols are implemented on top of a 10-node cluster of SQLite databases.
Experimental results show that ODES outperforms state-of-the-art schemes by orders of magnitude on the TPC-H benchmark and multiple real-world applications.

Our future work will focus on more efficient secret-sharing primitives that we hope will further help reduce the insertion and query time in database applications.
Another orthogonal research direction is to explore the feasibility of homomorphic encryption in order-preserving encryption in database systems.


\bibliographystyle{plain}
\bibliography{ref_new}

\end{document}